\documentclass[amsmath,amssymb,aps,pra,reprint,showpacs,twocolumn,floats,superscriptaddress,longbibliography]{revtex4-2}
    \usepackage[utf8]{inputenc}
    \usepackage[table,xcdraw]{xcolor}
    \usepackage{graphicx}
    \usepackage{amsmath, amsthm, amssymb, amsfonts}
    \usepackage{bbm}
    \usepackage{hyperref}
    \usepackage{footnote}
    \usepackage{siunitx}
    \usepackage{soul}
    \usepackage{capt-of}
    \usepackage{subfigure}
    \usepackage{microtype}
    \usepackage{physics}
    \usepackage{tikz, environ}
    \usepackage{multirow}
    \pdfoutput=1 
    \usepackage{easybmat}
    \usetikzlibrary{decorations.pathreplacing,calligraphy}
    
    \makeatletter
    \DeclareRobustCommand{\rvdots}{%
    \vbox{
        \baselineskip4\p@\lineskiplimit\z@
        \kern+\p@
        \hbox{.}\hbox{.}\hbox{.}
    }}
    \makeatother
    
    \newcommand\Tstrut{\rule{0pt}{2.6ex}}         
    \newcommand\Bstrut{\rule[-0.9ex]{0pt}{0pt}}   

    \renewcommand*{\binom}[2]{C\!\left(#1,#2\right)}
    
    \makeatletter
    \newcommand*\sdot{\mathpalette\sdot@{.5}}
    \newcommand*\sdot@[2]{\mathbin{\vcenter{\hbox{\scalebox{#2}{$\m@th#1\bullet$}}}}}
    \makeatother

    \newcommand{\LV}{\mathcal{L}}
    \newcommand{\LL}{\hat{c}}
    \newcommand{\HH}{\hat{H}}
    \newcommand{\dop}{\hat{\rho}}
    \newcommand{\sig}{\hat{\sigma}}
    \newcommand{\sigdag}{\hat{\sigma}^{\dagger}}
    \newcommand{\sigz}{\hat{\sigma}^z}
    \newcommand{\kpsi}{\ket{\psi}}

    \newcommand{\cbqty}[1]{\left\{#1\right\}}
    
    \newcommand{\bigentry}[2][\huge]{\begin{tabular}{@{}c@{}}#1$#2$\end{tabular}}

\begin{document}
\title{Scalable entanglement stabilization with modular reservoir engineering}
\author{E.~Doucet}
\email{emery\_doucet@student.uml.edu}
\affiliation{Department of Physics and Applied Physics, University of Massachusetts, Lowell, MA 01854, USA}
\author{L.~C.~G.~Govia}
\affiliation{IBM Quantum, IBM T.J. Watson Research Center, Yorktown Heights, NY 10598, USA}
\author{A.~Kamal}
\email{archana\_kamal@uml.edu}
\affiliation{Department of Physics and Applied Physics, University of Massachusetts, Lowell, MA 01854, USA}
\date{\today}
\begin{abstract}
Dissipation engineering is a powerful framework for quantum state preparation and autonomous error correction in few-qubit systems. In this work, we examine the scalability of this approach and give three criteria which any dissipative state stabilization protocol should satisfy to be truly scalable as the number of qubits grows.
Besides the requirement that it can be constructed in a resource-efficient manner from simple-to-engineer building blocks, a scalable protocol must also exhibit favorable scaling of the stabilization time with the increase in system size. 
We present a family of protocols which employ fixed-depth qubit-qubit interactions alongside engineered linear dissipation to stabilize an $N$-qubit W state. 
We find that a modular approach to dissipation engineering, with several overlapping few-qubit dissipators rather than a single $N$-qubit dissipator, is essential for our protocol to be scalable. 
With this approach, as the number of qubits increases our protocol exhibits low-degree polynomial scaling of the stabilization time and linear growth of the number of control drives in the best case. 
While the proposed protocol is most easily accessible with current state-of-the-art circuit-QED architectures, the modular dissipation engineering approach presented here can be readily adapted to other platforms and for stabilization of other interesting quantum states. 
\end{abstract}
\maketitle
%
%
\section{Introduction}
\label{sec:Intro}
%
Multipartite entanglement underpins almost every application of quantum information processing, ranging from quantum computing \cite{Raussendorf2001,Jozsa2003}, quantum sensing \cite{Cappellaro2017} to quantum communication \cite{Cirac1999,Gisin2007} and cryptography \cite{Pirandola2020}. Efficient preparation of entangled resource states is therefore a critical functionality and a key challenge for all quantum information architectures \cite{Preskill2012}. Typically, entangled states of some system -- say a set of qubits -- are prepared via the application of a sequence of entangling few-qubit unitary gates on the system which has been pre-initialized in some reference state. This approach is very flexible, general, and well-understood, however there are some drawbacks. The most obvious is that, once prepared, the entangled state will decay away due to the unavoidable effects of local decoherence. Thus, either the prepared state must be used quickly or some additional process must be included to mitigate the effects of decoherence, e.g. error correction \cite{Terhal2015,Georgescu2020}, dynamical decoupling \cite{Lidar2014,Suter2016}, or error mitigation \cite{Temme2017,Endo2018,OBrien2021} and suppression \cite{Koczor2021}.
\par
An alternative approach to such active methods of preparing and preserving quantum states is the dissipative state stabilization \cite{Verstraete2009, Harrington2022}, where the system is subjected to carefully engineered always-on interactions with an engineered environment leading to effective open dynamics on the system which cause it to relax to the desired entangled state. In addition to its inherent insensitivity to initialization and measurement errors, the primary attraction of this approach is that the prepared state is stable so long as the protocol is active, with high fidelities achievable even in the presence of spurious decoherence.
\par
Most dissipative state stabilization protocols proposed and implemented to date have focused on few-body entanglement stabilization, e.g. 2-qubit Bell states \cite{Doucet20,Brown21,Siddiqi2016,Shankar2013,Tureci2014,Liebfried2022,Schirmer2010,Motzoi2016,Reiter2012,Govia2022}, 3-qubit W states \cite{Cole2021}, bosonic cat states \cite{Clerk2018,Mirrahimi2015}, or small manifolds of states in bosonic systems \cite{Noh2022,Devoret2015,Mirrahimi2014}. Beyond this, the design and implementation of ``scalable'' dissipative stabilization protocols which can be extended to stabilize $N$-qubit entangled states persists as an open question. In fact, the notion of scalability itself in the context of dissipative state preparation and stabilization has remained largely unexplored until recently \cite{Sorensen16,Tureci16,Lukin2021,Petruccione2013}. 
\par
To clarify, let us first consider the quantum circuit model, where the problem of state preparation is broken into two parts. The first consists of resetting the system to some initial reference state $\ket{i}$ (e.g. all qubits in their ground states), usually through measurement and feedback or some other non-unitary process. This process usually has some small fixed cost which is negligible in large computations. The second and more complex part is to act some quantum circuit which implements a set of multi-qubit unitaries which rotate this initial state into the desired target state $\kpsi$. The study of the quantum circuit complexity is a very deep and well-developed field, but from the point of view of implementation the primary measure of complexity that matters is the number of gates required, or the circuit depth if the underlying hardware can run multiple gates in parallel. Many important entangled resource states can be prepared efficiently with quantum circuits, for example there exist circuits which prepare $N$-qubit GHZ states or W states with $\mathcal{O}(N)$ gates and $\mathcal{O}(\log N)$ depth \cite{Clement2019,Lloyd2021a}, and graph states can be prepared with constant-depth circuits with gate counts linear in the number of edges \cite{Lloyd2021b}. The existence of such resource-efficient circuits means that it is tractable to prepare these states, even as the number of qubits increases.
\par
In this work, inspired by the simple gate-counting notion of quantum circuit complexity, 
and extending the existing literature on scalable dissipative stabilization \cite{Tureci16,Sorensen16}, 
we lay out three criteria which should be satisfied for a dissipative stabilization protocol to be scalable. We then construct a family of protocols which prepare and stabilize $N$-qubit W states which satisfy all three proposed criteria for scalability. Crucially, we find engineering ``modular" dissipation, where several dissipators act on small, overlapping sets of qubits, is necessary for achieving scalable high-fidelity stabilization. It is worthwhile to note that the modular dissipation approach proposed here constitutes a significant departure from the conventional approach to dissipative stabilization, where typically a single ``global'' engineered dissipator acts on all qubits.
\par
The outline of the paper is as follows: in Sec.~\ref{sec:DissipativeStabilization} we present and motivate three key criteria which any dissipative stabilization protocol must satisfy to be scalable, and discuss some merits of the $N$-qubit W state as a target state for such a protocol. We then describe our design methodology for both the global and modular dissipator constructions in Sec.~\ref{sec:DissipatorConstruction}, followed by our system Hamiltonian design methodology in Sec.~\ref{sec:HamiltonianConstruction}. In Sec.~\ref{sec:Parameters}, we explore the parameter space for the proposed stabilization protocols, with a focus on both design freedom and minimality of number of interactions required. In Sec.~\ref{sec:Results}, we numerically study the performance of these protocols. The main results of this paper are presented in Fig.~\ref{fig:scaling} in this section, which shows the scaling of the stabilization time constant $\tau$ with the number of qubits $N$ for several variations of our protocol. Our results show that $\tau(N)$ exhibits low-degree polynomial scaling when employing the modular dissipator construction, whereas with a single global dissipator it exhibits exponential scaling. We conclude with a summary of our main results and offer perspectives in Sec.~\ref{ssec:Conclusion}. Additional mathematical details and simulations are included in Appendices~\ref{app:WDecay} through \ref{app:GammaDist}.
%
%
\section{Dissipative Stabilization}
\label{sec:DissipativeStabilization}
%
\subsection{Scalability}
\label{ssec:Scalability}
%
In contrast to the easily-counted discrete time decomposition of a quantum circuit, stabilization processes are built from always-on continuous processes. Scalability therefore is a property of the system and system-bath Hamiltonians. Generically, we can always decompose the system and system-bath Hamiltonians as,
\begin{subequations}
\begin{align}
    \HH_{\rm S} &= \sum_{j=1}^{n_{\rm S}} \lambda_j\hat{A}_{j,1}\dots\hat{A}_{j,m} \\
    \HH_{\rm SB} &= \sum_{j=1}^{n_{\rm SB}} g_j\hat{A}_{j,1}\dots\hat{A}_{j,p}\hat{B}_{j,1}\dots\hat{B}_{j,q}
    ,
    \label{eq:intro}
\end{align}
\end{subequations}
where each $\hat{A}_{j,k}$ ($\hat{B}_{j,k}$) is a local, linear operator acting on a single site of the system (bath) and identity everywhere else. E.g., Pauli operators for qubits or creation/annihilation operators for oscillators. Each term in $\HH_{\rm S/SB}$ consists of a product of some number of these local operators, we call this number the interaction depth. Each Hamiltonian has a maximum interaction depth, which we denote $m$ for $\HH_{\rm S}$ and $(p+q)$ for $\HH_{\rm SB}$, split between local system and bath operators. Once this decomposition has been identified, we propose that the following three criteria must be satisfied for a dissipative stabilization protocol to be scalable as the number of qubits $N$ increases:
\begin{enumerate}
    \item The maximum interaction depths $m$ and $(p+q)$ should remain bounded with increasing $N$.
    \item The number of engineered interactions $n_{\rm S}$ and $n_{\rm SB}$ should have some low-degree polynomial scaling with $N$.
    \item The time required for stabilization $\tau$ should have reasonable polynomial scaling with $N$ while the engineered coupling strengths $\lambda_j,g_j$ remain bounded.
\end{enumerate}
Presumably, in an experiment each term in $\HH_{\rm S/SB}$ will require some independent process to implement and optimize. The first criterion serves to ensure that these terms do not become arbitrarily complex. The second criterion essentially states that the resources required to engineer the Hamiltonian must scale efficiently if the resulting protocol is to be scalable. Note that we have chosen to require that the interaction depth be bounded, while the number of interactions may grow polynomially with $N$. This is a consequence of the experimental reality that the implementation of higher-depth interactions is typically more challenging than the simultaneous implementation of many lower-depth interactions. Previous works have explored hardware-efficient architectures for scalable dissipative stabilization in cavity-QED \cite{Tureci16} and trapped-ion  \cite{Sorensen16} platforms.
\par
While the first two criteria summarize hardware scalability, the last criterion ensures performance scalability by requiring that the resulting stabilization process itself completes in a reasonable time while the couplings remain bounded. In the presence of any decoherence or error processes, slow stabilization means reduced fidelity. A stabilization process which exhibits exponential slowdown with $N$ will not scale to large $N$, as the decoherence processes will rapidly overwhelm the stabilization mechanism unless the various coupling and engineered decay rates are rapidly increased to compensate, which will eventually cause problems itself. For instance, a scalable GHZ stabilization protocol employing four-level ``atoms'' was presented in \cite{Sorensen16}, where the authors found that power broadening due to strong driving limits the stabilization time to a polynomial scaling of $\tau \sim\mathcal{O}(N^{3/2}\log N)$. 
\par
In this work, we construct a family of dissipative stabilization protocols which prepare $N$-qubit W states, which satisfy all three of our criteria for scalability. We perform a detailed numerical investigation of performance scaling of these protocols, and compare it for two distinct classes of dissipation engineering. Employing fixed-depth interactions ($p=q=1, m\le3$), our best protocol exhibits linear resource scaling $(n_{\rm s}, n_{\rm SB})$ with the number of qubits $N$, and a stabilization time that appears to grow only quadratically with $N$. 
%
\subsection{Target state choice}
\label{ssec:TargetState}
%
The goal of dissipative state stabilization is to engineer effective open system dynamics which cause a system to eventually relax into a specific target state, regardless of the initial state $\dop(0)$. Here we will focus on stabilizing the $N$-qubit W state, defined as
\begin{align}
    \ket{W^N} 
    &= \frac{1}{\sqrt{N}}\sum_{j=1}^N \sigdag_j\ket{0^{\otimes N}} \nonumber\\
    &= \frac{1}{\sqrt{N}}\big(
        \ket{10\cdots0} + \dots + \ket{0\cdots01}
    \big),
    \label{eqn:WState}
\end{align}
which is the simplest type of Dicke state with Hamming weight one. This general class of multipartite entangled states finds several uses in the context of quantum metrology \cite{Ouyang2022}, multiparty quantum communication \cite{Prevdel2009} and combinatorial optimization algorithms \cite{Hadfield2019}. Given their robustness to particle loss and depoloarization errors \cite{Stockton2003,Barnes2015}, it has been shown that W-states can lead to stronger nonclassicality and higher rates of entanglement distillation than GHZ states \cite{Sen2003}. 
\par
Our choice of target state $\ket{W^N}$ is directly motivated by the notion of scalability given in the previous section. Specifically, we seek to construct dissipative stabilization protocols which employ only linear engineered dissipation \cite{Ticozzi2014}, i.e. $p = 1$ in Eq.~(\ref{eq:intro}). Such linear dissipation is quite easy to implement experimentally since it does not need higher-order nonlinear interactions, making the proposed protocols accessible for near-term applications. As discussed in more detail in the next section on dissipator construction, this choice of dissipation is fairly restrictive as to which interesting entangled states can be stabilized. For example, this precludes the stabilization of the $N$-qubit GHZ state, a critical resource state for quantum error correction with fundamentally distinct entanglement structure from $\ket{W^{N}}$.
\par
Another consideration guiding our choice of target state is the robustness of $\ket{W^N}$  against uncontrolled sources of dissipation or decoherence, such as local qubit relaxation or dephasing. In general, the target state $\kpsi$ is not stationary under the action of these uncontrolled sources of dissipation; 
nonetheless, provided we ensure that the protocol is operated in the ``dissipation engineering'' regime, i.e. where the engineered dissipation rates are much larger than the effective decoherence rates, then the long-time steady state of the system can still be unique and have a high fidelity to the desired target state $\kpsi$. In this regime, the stabilization protocol is well-approximated as a competition between the stabilization rate and the effective decay rate of $\kpsi$, with the steady-state error being the ratio of these rates. As shown in Appendix~\ref{app:WDecay}, the decay rate of $\ket*{W^N}$,
\begin{align}
    \gamma_\downarrow^{W} &\approx \Bar{\gamma}^{-} + 4\frac{N-1}{N}\Bar{\gamma}^{z}
    ,
\end{align}
where $\Bar{\gamma}^{-}$ ($\Bar{\gamma}^{z}$) is the average single qubit decay (dephasing) rate, remains bounded with $N$ and in fact saturates to $\Bar{\gamma}^{-} + 4\Bar{\gamma}^{z}$ in the large-$N$ limit. This property is crucial for scalable dissipative stabilization of these states, as $\gamma_\downarrow^{W}$ is the rate which the stabilization rate $\Gamma_{\uparrow} \equiv \tau^{-1}$ must overwhelm to have a small steady-state error, $\varepsilon_\infty\approx\gamma_\downarrow/\Gamma_{\uparrow}$. This reduces the need for stronger couplings or lower decoherence rates as the system increases in size, as long as $\Gamma_{\uparrow}$ decreases only weakly as $N$ grows. The $N$-qubit GHZ state provides a counterpoint, since $\gamma_\downarrow^{\rm GHZ} \approx \frac{N}{2}\left(\gamma^- + 2\gamma^z\right)$ increases without bound with $N$. It is therefore increasingly difficult to reach the dissipation engineering regime when stabilizing large GHZ states, which conflicts with our third criteria for a scalable stabilization protocol presented in Sec.~\ref{ssec:Scalability}.
\par
Further, we will not consider approximate stabilization protocols, which rely on stabilizing a pure state $\ket{\psi'} = \mathcal{N}\pqty{\ket{\psi} + \epsilon\ket{\xi}}$ which has a large overlap with the desired target state $\ket{\psi}$. Several works have explored such protocols for stabilizing Bell states, as the freedom available in choosing $\ket{\xi}$ can translate into simpler interactions; for instance, only local drives and linear dissipation \cite{Motzoi2016,Govia2022}. However, these protocols always have some minimum intrinsic steady-state error, $\sim|\epsilon|^{2}$, which invariably leads to a tradeoff between the stabilization time ($\tau$) and achievable steady-state error ($\varepsilon_{\infty}$)\cite{Doucet20}. We choose to exclude such approximate stabilization protocols for two reasons: firstly, as the number of qubits grows, it becomes increasingly non-trivial to choose what ``nearby'' state $\ket*{\psi'}$ may be optimal. Secondly, by considering protocols whose steady state is \emph{exactly} the entangled state $\ket{\psi}$, we do not have to concern ourselves with the scaling of the intrinsic error with system size and system-reservoir coupling strengths. 
%
%
%
\section{Dissipator construction}
\label{sec:DissipatorConstruction}
%
%
To analyze our stabilization protocols, we model the engineered $N$-qubit dynamics using a Markovian master equation,
\begin{align}
    \dv{\dop(t)}{t} = 
    \LV \dop(t) \equiv 
    -i\comm{\lambda\HH_{\rm S}}{\dop(t)} 
    + \sum_{j=1}^m \Gamma_j \mathcal{D}\left[\LL_j\right] \dop(t)
    ,
    \label{eqn:GeneralME}
\end{align}
where the superoperators $\mathcal{D}\left[\LL_j\right]\sdot = \LL_j\sdot\LL_j^\dagger - (1/2)\{\LL_j^\dagger\LL_j, \sdot\}$ are Lindblad dissipators with arbitrary engineered jump operators $\LL_j$. Carefully choosing the forms and strengths of the interactions in $\HH_{\rm SB}$ selects the jump operators $\LL_j$ and controls the effective decay rates $\Gamma_j$ respectively. In this work, we will not concern ourselves with the details of the microscopic derivation of Eq.~\eqref{eqn:GeneralME}. Assuming $\HH_{\rm SB}$ is sufficiently weak, one can follow standard methods \cite{Carmichael2} to construct a Lindblad master equation for the reduced system of qubits.
\par
For the purposes of the modular approach proposed here, we use one or more linear engineered dissipators of the form
\begin{align}
    \mathcal{D}\bqty{\LL_{E}} 
    \equiv \mathcal{D}\bqty{\sum_{j\in E}\pqty{r_j[E] \sig_j + s_j[E]\sigdag_j}}
    ,
    \label{eqn:QuasiLocalDissipation}
\end{align}
where $E$ denotes a set of qubit indices indicating on which qubits the associated jump operator $\LL_E$ acts. The coefficients $r_j[E],s_j[E]\in\mathbb{C}$ can be chosen by tuning the coefficients $\{g_{j}\}$ in $\HH_{\rm SB}$ [Eq.~(\ref{eq:intro})]. We refer to a dissipator acting on $k$ qubits (i.e. $|E| \equiv k\le N$) as a ``width-$k$'' dissipator. If we represent each qubit as a node of a hypergraph, then we can interpret the set $E$ corresponding to each dissipator as defining a hyperedge connecting the relevant qubits. This picture will be useful for representing and understanding several features of our dissipator construction moving forward. 
\par
In principle, for an arbitrary pure state $\kpsi$, one can always construct a Lindblad-form master equation whose unique steady state is that state \cite{Zoller08}. This requires that the state $\kpsi$ be made a \emph{dark state} of the dynamics, meaning it must both be an eigenstate of $\HH_{\rm S}$ and be a null state of every $\LL_j$ (i.e. $\LL_j\kpsi = 0 ~\forall j$). Together, these two criteria ensure that $\dyad{\psi}{\psi}$ is a stationary state of the master equation \eqref{eqn:GeneralME}. We must additionally require that $\kpsi$ is the only dark state of the dynamics, otherwise there would be some other stationary state $\ket{\phi}$ which does not eventually relax into the target state. This second requirement can be broadened to requiring that there is no subset $S$ of the eigenstates of $\HH_{\rm S}$ which is closed under the action of all the jump operators $\LL_j$ and which does not include $\kpsi$. If there were, then the projection of the initial state onto $S$, i.e. $\Tr\!\left.[\hat{\Pi}_S\dop(0)]\right.$, would be preserved and any state in that subspace would again not relax into the target state. Provided $\kpsi$ is the only dark state of the dynamics and there is no closed subspace as just described, then the long time steady state of the dynamics is
\begin{align}
    \dop(t\to\infty) = \dyad{\psi}{\psi}
    ,
\end{align}
independent of the initial state $\dop(0)$. 
\par
The stationary states of a dissipator described by a jump operator $\LL_E$ are those states which are contained in the kernel (nullspace) of $\LL_E$, denoted by $\ker(\LL_E)$. Under the action of $m$ independent dissipators of the form shown in Eq.~(\ref{eqn:QuasiLocalDissipation}), we must have
\begin{align}
    \ket{\psi} \in \bigcap_{j=1}^m \ker\!\pqty{\LL_{E_j}}
    ,
    \label{eqn:KerIntersect}
\end{align}
for $\ket{\psi}$ to be a steady state of the joint dynamics.
\par
As mentioned earlier, the linear nature of each jump operator on the right-hand side of Eq.~(\ref{eqn:QuasiLocalDissipation}) strongly restricts the form of the multipartite entangled states that can appear as null states of the joint dissipative dynamics. For example, the $N$-qubit GHZ state $\pqty{\ket*{0^{\otimes N}} + \ket*{1^{\otimes N}}}/\sqrt{2}$ cannot be in any $\ker(\LL_E)$ unless every coefficient describing the jump operator is zero, in which case $\LL_E = 0$ and there is no dissipation at all. Clearly, it is not possible to stabilize GHZ states with linear dissipation alone, and in fact the same conclusion can be drawn for many other $N$-qubit entangled states, for example the four-qubit cluster state $\ket{\mathcal{C}_{4}} = (\ket{0000} + \ket{0011} + \ket{1100} - \ket{1111})/2$. 
\par
Fortunately, since $\ket*{W^N}$ is a single-excitation state, it is possible for each $\LL_E$ to have $\ket*{W^N}$ in its kernel. To show this, we first consider a single engineered dissipator of width-$N$, i.e. $E=\cbqty{1,\dots,N}$ indicating that this dissipator acts on all qubits together. To construct such a jump operator $\LL_E$, with $\ket*{W^N} \in \ker(\LL_E)$, we must require that
\begin{align}
    \LL_E \ket*{W^N} 
    &= \sum_{j=1}^N\pqty{r_j[E] \sig_j + s_j[E]\sigdag_j}\ket*{W^N} \nonumber\\
    &= \sum_{j=1}^N r_j[E]\ket*{0^{\otimes N}} \nonumber\\
    &\quad+ \sum_{j=1}^{N-1}\sum_{k=j+1}^N \pqty{s_j[E] + s_k[E]}\ket{jk} \nonumber\\
    &= 0
    ,
    \label{eq:nullstate}
\end{align}
where we have used the shorthand notation ${\ket{jk}\equiv\sigdag_j\sigdag_k\ket*{0^{\otimes N}}}$ to represent arbitrary two-excitation product states. For Eq.~(\ref{eq:nullstate}) to satisfied, we must choose the coefficients such that, 
\begin{alignat}{3}
    \sum_{j=1}^N r_j[E] = 0 
    & \qquad \textrm{and} \qquad &
    s_j[E] = 0 ~~ \forall j
    ,
    \label{eqn:LLConstraints}
\end{alignat}
leading to a jump operator which is purely a linear combination of single-qubit lowering operators. For an $N$-qubit jump operator $\LL_E$, we find that ${\dim[\ker(\LL_E)] = \binom{N}{\lfloor N/2\rfloor}}$, where we use $\binom{N}{k}$ to represent the binomial coefficient. By construction $\ket*{W^N}\in\ker(\LL_E)$, and due to the fact that only lowering operators enter the jump operator, we also have $\ket*{0^{\otimes N}}\in\ker(\LL_E)$. The structure of the remaining states depends on $N$: for example, when $N=3$ the only other state stationary under $\LL_E$ is a single-excitation state, whereas for $N=4$ there are both one- and two-excitation states in $\ker(\LL_E)$. 
\begin{figure*}[t!]
    \centering
    \includegraphics[width=\textwidth]{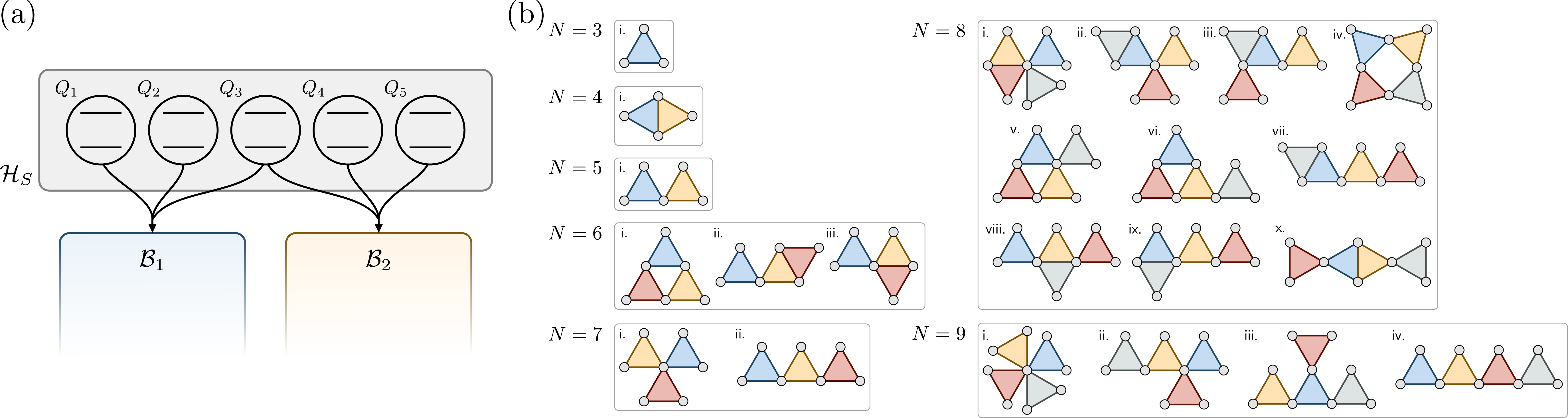}
    \caption{(a) Schematic representation of one possible `modular' dissipator configuration, consisting of five qubits coupled to two independent engineered baths. The coupling of qubits $\cbqty{1,2,3}$ to $\mathcal{B}_1$ engineers a width-$3$ dissipator of the form shown in Eq.~\eqref{eqn:QuasiLocalDissipation} with $E_1 = \{1,2,3\}$. Similarly, the coupling to $\mathcal{B}_2$ engineers a second dissipator with $E_2 = \{3,4,5\}$. 
    (b) Illustration of different connected configurations of the smallest number of width-3 dissipators for $3 \le N \le 9$. Each gray node represents a qubit, and each colored triangle represents a hyperedge connecting the nodes. This, in turn, represents an independent engineered dissipator acting on the relevant three qubits. For example, the sole configuration shown for $N=5$ corresponds to the schematic shown in (a).}
    \label{fig:diagrams}
\end{figure*}
\par
Note that for $N=2$ qubits, a linear dissipator of the form shown in Eq.~\eqref{eqn:QuasiLocalDissipation} with coefficients chosen subject to the constraints in Eq.~\ref{eqn:LLConstraints} will not lead to a stabilization protocol for a $2$-qubit ``W state'' -- really a Bell state. We have explored this issue and presented a workaround employing qudits in a previous work on Bell state stabilization \cite{Brown21}, which can be briefly summarized by noting that for $N=2$, if $\LL_E\ket*{W^2}=0$, then $\LL_E^\dagger\ket*{W^2}=0$ as well. This indicates that while $\ket*{W^2}$ is a null state of the width-2 dissipation, such a linear dissipator cannot couple any other states into $\ket*{W^2}$ and hence cannot prepare it. This issue can be circumvented by using more complicated engineered dissipation, for instance, one that involves qudits \cite{Brown21} or combines longitudinal and transverse interactions \cite{Doucet20,Leghtas2013,Shankar2013}. In the interest of simplicity of implementation, in this work we choose to work with qubits and purely transverse interactions, and thus will consider dissipators of minimum width three ($k\ge3$).
\par
In conjunction with a Hamiltonian $\HH_{\rm S}$, which breaks the degeneracy between $\ket{W^N}$ and the other states in $\ker(\LL_E)$ -- most importantly $\ket*{0^{\otimes N}}$ -- our engineered $N$-qubit dissipation leads to a protocol that stabilizes $\ket{W^N}$. This approach is conceptually similar to previous works by us and others where a single dissipator acts on all qubits, and a Hamiltonian interaction is employed to break any degeneracies in the steady state while providing enough connectivity that any initial state flows into the target state. We call this approach `global' dissipation engineering in order to contrast it with the `modular' dissipation engineering approach which is the primary focus of this paper. 
\par
To motivate a modular protocol for stabilizing $\ket*{W^N}$, it is instructive to rewrite the corresponding wavefunction in the following form, where instead of the symmetric form shown in Eq.~(\ref{eqn:WState}) we can privilege the first $k$ qubits to write,
\begin{align}
    \ket{W^N} &= 
    \sqrt{\frac{k}{N}}\ket{W^k}\ket{0^{\otimes N-k}}
    +
    \sqrt{\frac{N-k}{N}}\ket{0^{\otimes k}}\ket{W^{N-k}}. 
    \label{eqn:WStatemod}
\end{align}
This shows that $\ket*{W^{N}}$ may be understood as a superposition of the tensor product of $\ket*{W^{k}}$ on the first $k$ qubits and the ground state on the remaining qubits, with the complementary state consisting of the ground state on the first $k$ qubits and the $\ket*{W^{N-k}}$ on the remaining qubits. We know that any width-$k$ $\LL_{\cbqty{1,\dots,k}}$ whose coefficients satisfy Eq.~\eqref{eqn:LLConstraints} will include both $\ket*{W^k}$ and $\ket*{0^{\otimes k}}$ in its kernel, and thus per Eq.~\eqref{eqn:WStatemod} $\LL_{\cbqty{1,\dots,k}}\ket*{W^N} = 0$. Of course, only the state of the first $k$ qubits will be constrained and so \emph{any} state of the form $\alpha\ket*{0^{\otimes k}}\ket{\phi} + \beta\ket*{W^k}\ket{\chi}$ will be a stationary state of this width-$k$ jump operator. A similar approach was taken in \cite{Lukin2021} for stabilizing an Affleck-Kennedy-Lieb-Tasaki (AKLT) ground state, employing multiple dissipators acting on overlapping pairs of qubits. 
%
\par
In order to constrain the dissipative dynamics and have fewer stationary states, we can add a second width-$k$ dissipator which acts on a different subset of the $N$ qubits, rather than just the first $k$. The set of stationary states of the dynamics under both dissipators is the intersection of the kernels of the jump operators describing each dissipator, as shown in Eq.~(\eqref{eqn:KerIntersect}). Thus, by introducing additional dissipators, we progressively reduce the dimension of the joint nullspace of all the jump operators. Since we necessarily have that $\ker(\LL_{E_j})\supset\{\ket*{W^{|E_j|}}, \ket*{0^{\otimes|E_j|}}\}$, $\ket*{W^{N}}$ can never be the only steady state of the joint dissipative dynamics; instead we always have 
\begin{align}
    \bigcap_{j=1}^m \ker\!\pqty{\LL_{E_j}} \supseteq \{\ket*{W^{N}}, \ket*{0^{\otimes N}}\}
    ,
    \label{eq:disskernel}
\end{align}
where number and form of additional stationary states depend on the connectivity of the dissipator hypergraph and the particular coefficients chosen for each dissipator. Figure~\ref{fig:diagrams}(a) shows a schematic representation of this modular approach to dissipation engineering, illustrating a minimal construction for $N=5$ qubits with two width-$3$ jump operators, $\LL_{\cbqty{1,2,3}}$ and $\LL_{\cbqty{3,4,5}}$, acting on the first three and last three qubits respectively. Jointly, their nullspace consists of the states $\{\ket*{W^5}, \ket*{00000}\}$ and a few additional irrelevant states, similar to what was found with a width-$5$ global dissipator. 
\par
If we represent the dissipative couplings with a hypergraph where each node represents a qubit and each hyperedge connecting $k$ nodes represents a width-$k$ dissipator acting on those qubits, then we find that connected hypergraphs lead to engineered dissipation which excludes spurious W-like stationary states (see Appendix~\ref{app:OverlappingDissipators} for details). In order to optimize the resource efficiency of our protocol and to simplify the Hamiltonian design, we will focus on minimal configurations, corresponding to connected hypergraphs with the fewest possible edges. With width-$3$ dissipators (the smallest width we consider in this work) we require at least $\lfloor N/2\rfloor$ dissipators to construct a connected hypergraph. Figure~(\ref{fig:diagrams}b) shows the hypergraphs representing all possible minimal configurations for up to 9 qubits.
\par
We now have two conceptually distinct methodologies for constructing the dissipative part of a stabilization protocol, both leading to an overall nullspace spanned by $\{\ket*{00\cdots}, \ket*{W^N}\}$ and possibly some additional irrelevant states which we find do not cause any issues. As we explore the stabilization performance of these different forms of our protocol in Sec.~\ref{sec:Results}, we will find that the overlapping construction leads to significantly better performance scaling with the number of qubits $N$. Additionally, we expect that the overlapping dissipator construction will lead to easier implementations in near-term experiments as it requires engineering dissipation on only a few qubits at a time rather than all $N$ at once.
%
\section{Hamiltonian construction}
\label{sec:HamiltonianConstruction}
%
Having described the dissipator construction, we now turn our attention to the design of the system Hamiltonian, $\HH_{\rm S}$. 
As discussed in the previous section, the purpose of the Hamiltonian is to ensure that $\ket*{W^{N}}$ is the only state stationary under the dissipation which is \emph{simultaneously} an eigenstate of the Hamiltonian. Most importantly, in view of Eq.~(\ref{eq:disskernel}), the state $\ket*{0^{\otimes N}}$, should not remain invariant under the action of $\HH_{\rm S}$ in order for $\ket*{W^N}$ to be the unique steady state of the driven-dissipative dynamics. We restrict $\HH_{\rm S}$ to include only transverse (unconditional) interactions, motivated by the comparative ease with which such interactions can be implemented with currently accessible experimental techniques.
\par
It is clear that there must be at least one term in $\HH_{\rm S}$ consisting solely of a product of raising operators $\sigdag_i\dots\sigdag_k$ so that $\ket*{0^{\otimes N}}$ is not an eigenstate. As for the other states which are stationary under the dissipation, we find that we may ignore them when designing $\HH_{\rm S}$, since these extra stationary states are never ``coincidentally'' eigenstates of $\HH_{\rm S}$ and, therefore, do not meaningfully affect the resultant stabilization protocol. In short, unless we explicitly design the $\HH_{\rm S}$ to include these states as eigenstates, it is very unlikely that there is any eigenstate of $\HH_{\rm S}$ which is in the space spanned by the states stationary under the dissipation. We discuss this statement in more detail in Appendix~\ref{app:OtherStates}; however for the remainder of this work we will ignore these other states.
\par
In keeping with the notion of scalability presented in Sec.~\ref{ssec:Scalability}, we would like to minimize the interaction depth of each term in $\HH_{\rm S}$. We can see that a system Hamiltonian consisting entirely of linear drive terms, $\HH_{\rm S}^{(1)}$, will not suffice since
\begin{align}
    \HH_{\rm S}^{(1)}&\ket{W^N} 
    = \sum_{j=1}^N \pqty{a_j\sigdag_j + h.c.}\ket{W^N} \nonumber\\
    &= \frac{1}{\sqrt{N}}\sum_{j=1}^N\sum_{k=1}^N \pqty{a_j\sigdag_j\sigdag_k + a_j^*\sig_j\sigdag_k}\ket{0^{\otimes N}} \nonumber\\
    &= \frac{1}{\sqrt{N}}\sum_{j=1}^{N-1}\sum_{k=j+1}^N \pqty{a_j + a_k}\ket{jk} + \frac{1}{\sqrt{N}}\sum_{j=1}^N a_j^* \ket{0^{\otimes N}}
    .
\end{align}
The state $\ket{W^N}$ is only an eigenstate of $\HH_{\rm S}^{(1)}$ if the equation above is identically zero, which would require that the coefficients $a_j \in \mathbb{C}$ satisfy
\begin{alignat}{3}
    a_j + a_k &= 0 ~~\forall j<k
    & \qquad \textrm{and} \qquad &
    \sum_{j=1}^N a_j^* &= 0.
\end{alignat}
This is not possible unless we choose \emph{all} $a_j=0$, which implies $\HH_{\rm S}^{(1)} = 0$. Thus, we must include multi-qubit interactions in $\HH_{\rm S}$.
\par
Allowing at most bilinear interactions, the most general Hamiltonian one can consider is
\begin{align}
    \HH_{\rm S}^{(2)} = \sum_{j=1}^N a_j \sigdag_j
    + \sum_{j=1}^{N-1}\sum_{k=j+1}^{N} \pqty{c_{jk}\sigdag_j\sigdag_k + d_{jk}\sigdag_j\sig_k}
    +h.c.,
    \label{eqn:BilinearH}
\end{align}
which includes both hopping and two-qubit excitation interactions, in addition to linear driving terms. Note that when acting on a state with a well-defined excitation number (like $\ket*{W^N}$ or $\ket*{0^{\otimes N}}$) the hopping interactions conserve the excitation number, the linear drives change it by one and the two-qubit excitation interactions by two. Since each interaction type couples $\ket*{W^N}$ to a different set of states, the constraints on their coefficients are independent. 
\par
%
We are forced to set ${a_j = 0}$ by the same logic as followed when analyzing $\HH_{\rm S}^{(1)}$. 
For the hopping interactions, we are free to choose any parameters which support the $\ket{W^N}$ as an eigenstate; however, no choice leads to coupling the joint ground state $\ket*{0^{\otimes N}}$ to any other state. Therefore, they are not a useful addition to the Hamiltonian in terms of ensuring that $\ket*{0^{\otimes N}}$ is not an eigenstate of $\HH_{\rm S}$. 
\par
\begin{figure}[t!]
    \centering
    \includegraphics[width=0.93\columnwidth]{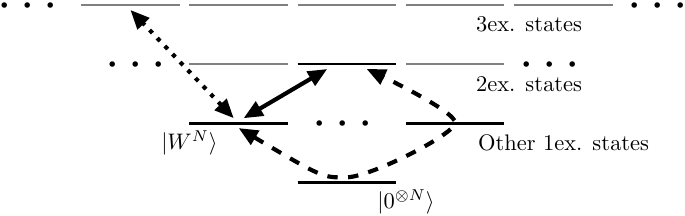}
    \caption{Schematic illustration of how $\ket*{W^N}$ is coupled into the manifolds of states with different excitation numbers by $\sigdag_j$ (solid), $\sigdag_j\sigdag_k$ (dotted), and $\sig_j\sigdag_k\sigdag_l$ (dashed) type Hamiltonian interactions. The linear and generalized hopping interactions can interfere destructively with one another, allowing $\ket{W^N}$ to be a null state of $\HH_{\rm S}$ in Eq.~(\ref{eqn:FinalH}).}
    \label{fig:exladder}
\end{figure}

This leaves only the two-qubit excitation interactions. Since $\ket*{W^N}$ is a one-excitation state, this interaction couples it into the space of three-excitation states, and so the only possible eigenvalue $\ket*{W^N}$ could have is zero. The constraints on the coefficients are then
\begin{align}
    c_{jk} + c_{jl} + c_{kl} = 0 ~~ \forall j<k<l
    .
\end{align}
While it is possible to find coefficients satisfying these coefficients when $N=3$ or $N=4$, those are the only two cases where a non-zero solution exists. When $N\ge5$, the $\ket*{W^N}$ cannot be an eigenstate of two-qubit excitation interactions. In Appendix~\ref{app:Bilinear} we prove this fact, but there is a straightforward heuristic argument as to why this should be expected. Each of the constraints on the coefficients comes from the requirement that $\mel{jkl}{\HH_{\rm S}^{(2)}}{W^N} = 0$, where $\ket{jkl} \equiv \sigdag_j\sigdag_k\sigdag_l\ket*{0^{\otimes N}}$ is an arbitrary three-excitation state. There are $\binom{N}{3}\sim\mathcal{O}(N^3)$ such states, whereas there are only $\binom{N}{2}\sim\mathcal{O}(N^2)$ coefficients, one for each pair of qubits. For $N\ge 5$, because there are at least as many constraints as there are parameters a solution may not (and, in fact, does not) exist. One could apply the same logic to analyze linear driving Hamiltonian $\HH_{\rm S}^{(1)}$ or any other Hamiltonian consisting of only qubit raising operators to prove that such interactions cannot support $\ket*{W^N}$ as an eigenstate for all $N$. Figure~\ref{fig:exladder} illustrates this issue, showing how $\HH_{\rm S}^{(1)}$ and $\HH_{\rm S}^{(2)}$ couple $\ket*{W^N}$ into the larger manifolds of two- and three-excitation states, respectively.
\par
If we expand our candidate interactions to include general trilinear interactions, we find a family of Hamiltonians that sidestep this issue. The most generic form of such a Hamiltonian is
\begin{align}
    \HH_{\rm S}^{(3)} &= 
    \sum_{j=1}^N a_j \sigdag_j
    + \sum_{j=1}^{N-1}\sum_{k=j+1}^{N} \pqty{c_{jk}\sigdag_j\sigdag_k + d_{jk}\sigdag_j\sig_k} \nonumber\\
    &\quad + \sum_{j=1}^{N-2}\sum_{k=j+1}^{N-1}\sum_{l=k+1}^{N} c_{jkl}\sigdag_j\sigdag_k\sigdag_l \nonumber\\
    &\quad + \sum_{j=1}^{N}\sum_{\substack{k,l\ne j\\k<l}}f_{j,kl}\sig_j\sigdag_k\sigdag_l
    +h.c.
\end{align}
As before we may separate the terms of $\HH_{\rm S}^{(3)}$ based on the number of excitations they add or remove. The crucial difference in this case is that there are two different types of terms which add or remove one excitation: both the linear driving terms and the generalized hopping terms, which de-excite one (two) qubits and excite two (one) others. These two types of interactions together can support $\ket*{W^N}$ as an eigenstate for arbitrary $N$. This leaves us with the final form of the Hamiltonian we use for stabilization:
\begin{align}
    \HH_{\rm S} &= 
    \sum_{j=1}^N a_j \sigdag_j
    + \sum_{j=1}^{N}\sum_{\substack{k,l\ne j\\k<l}}f_{j,kl}\sig_j\sigdag_k\sigdag_l
    +h.c.
    \label{eqn:FinalH}
\end{align}
\par
Figure~\ref{fig:exladder} illustrates how the combination of linear driving and these generalized hopping interactions can destructively interfere with one another, suppressing the coupling out of $\ket*{W^N}$. Specifically, $\HH_{\rm S}$ couples the $\ket*{W^N}$ into the zero- and two-excitation subspaces, and so $\ket*{W^N}$ will be an eigenstate with eigenvalue zero if and only if $\mel*{0^{\otimes N}}{\HH_{\rm S}}{W^N} = \mel*{kl}{\HH_{\rm S}}{W^N} = 0$. Computing these matrix elements leads to the following set of constraints on the Hamiltonian parameters,
\begin{subequations}
\begin{align}
    a_k + a_l + \sum_{j\ne k,l} f_{j,kl} &= 0 ~~ \forall k<l
    \label{eqn:HCubConstraintsA}
    \\
    \sum_{j=1}^N a_j^* &= 0
    .
\end{align}
    \label{eqn:HCubConstraints}%
\end{subequations}
This is an underconstrained problem, with $1+\binom{N}{2}\sim\mathcal{O}(N^2)$ constraints on $N+N\binom{N-1}{2}\sim\mathcal{O}(N^3)$ coefficients. Therefore, there exists an enormous variety of solutions for any $N$.
\par
Examining the constraints in Eq.~\eqref{eqn:HCubConstraintsA}, we notice that only the linear drive coefficients $\cbqty{a_j}$ will be present in multiple constraint equations. The trilinear drive coefficients $f_{j,kl}$ only contribute to the constraint coming from the $\mel{kl}{H^{(3)}}{W^N}$ matrix element. This greatly simplifies the process of choosing the coefficients in the Hamiltonian, which can be summarized as
\begin{enumerate}
    \item Choose a set of complex coefficients $\cbqty{a_j}$ which sum to zero, where at least one $a_j\ne0$. Simple examples include the $N$-th roots of unity, $a_j = \exp\pqty{\frac{2\pi i}{N}j}$, or the choice $a_1 = -a_2 = 1, a_{j>2}=0$.
    \item For each $(k,l)$, choose a set of complex coefficients $\cbqty{f_{j,lk}}$ which sum to $-(a_k+a_l)$. If $(a_k+a_l)=0$, we may choose these coefficients to be zero. If not, simple choices include distributing the sum equally among each coefficient $f_{j,lk}=(a_k+a_l)/(N-2)$, or setting one coefficient to $-(a_k+a_l)$ and the remainder to zero.
\end{enumerate}
\section{Parameter Design}
\label{sec:Parameters}
%
A complete description of a protocol for stabilizing $\ket*{W^N}$ consists of a set of coefficients for the linear ($\{a_j\}$) and trilinear ($\{f_{j,kl}\}$) interactions in the Hamiltonian, a set of edges ($\{E_j\}$) indicating the structure of the dissipation and coefficients ($\{r_k[E_j]\}$) for each jump operator. Once chosen, the stabilization dynamics are described with the master equation given in  Eq.~\eqref{eqn:GeneralME}. 
%
\subsection{Dissipators}
%
The modular nature of our protocol means we have the freedom to engineer the dissipation separately from the Hamiltonian. We have presented two distinct approaches to dissipation engineering, the first consisting of a single width-$N$ (`global') dissipator which couples to all $N$ qubits simultaneously, represented with a single hyperedge $E = \{1,\dots,N\}$. Note that such global dissipators lead to maximally-connected hypergraphs, as shown on the top left of Fig~\ref{fig:scaling} for various $N$. The coefficients which enter the jump operator for this dissipator are underconstrained. The specific choice of coefficients $\{r_j[E]\}$ does have an effect on the performance of the protocol, however as we will discuss in Sec.~\ref{sec:Results}b we find that most choices lead to similar performance. As a representative example, we choose the coefficients
\begin{align}
    r_j[E] &= \exp(2\pi i\frac{j-1}{N}) + (1-\delta^N_j)\exp(2\pi i\frac{j-1}{N-1}) 
    ,
    \label{eqn:CoeffsLOne}
\end{align}
formed using a combination of the $N$'th and $(N-1)$'th roots of unity. We find this combination does a better job of excluding other W-like states than using the $N$'th roots of unity alone.
\par
%
The second approach to dissipation engineering we presented is the `modular' approach, employing several few-qubit dissipators.
There is significant freedom in the design of such modular dissipator configurations, therefore we typically focus on the most resource-efficient configurations. To this end, we fix the width of each dissipator at $k=3$ -- the smallest allowed with linear dissipation -- and focus on configurations which are both connected and minimal. 
\par
For $k=3$, this means we consider configurations with $\lfloor N/2\rfloor$ dissipators, the minimum required to have a connected hypergraph as shown in Appendix \ref{app:OverlappingDissipators}. For $3\le N \le9$ these are precisely the same dissipator configurations that are illustrated in Fig.~\ref{fig:diagrams}(b). For the specific case of $N=5$ qubits, we explore the full landscape of possible width-3 dissipator configurations, minimal and non-minimal, in Sec.\ref{ssec:CompareConfigs}. For other $N$, we desire a simple family of configurations which are easy to extend to more qubits. To this end, we choose for odd $N$ the ``chain'' configuration illustrated in the top center of Fig~\ref{fig:scaling} and described by the set of hyperedges $\{E_1,\dots,E_{(N-1)/2}\}$ where
\begin{align}
    E_1 &= \{1, 2, 3\} \nonumber\\ 
    E_2 &= \{3, 4, 5\} \nonumber\\
        &\cdots \nonumber\\ 
    E_{(N-1)/2} &= \{N-2, N-1, N\}
    .
\end{align}
For even $N$ we use the ``ring'' configuration illustrated in the top right of Fig~\ref{fig:scaling} and described by the hyperedges $\{E_1,\dots,E_{N/2}\}$ where
\begin{align}
    E_1 &= \{1, 2, 3\} \nonumber\\ 
    E_2 &= \{3, 4, 5\} \nonumber\\
        &\cdots \nonumber\\ 
    E_{N/2} &= \{N-1, N, 1\}
    .
\end{align}
\par
Each of the width-3 dissipators in these configurations has three complex coefficients $\{r_1[E],r_2[E],r_3[E]\}$, which can be constrained by evaluating the action of the engineered jump operators on the three states $\{\ket*{W^3}, \ket*{\tilde{W}^3_+}, \ket*{\tilde{W}^3_-}\}$, defined as
\begin{subequations}
\begin{align}
    \ket*{\tilde{W}^3_+} &= \frac{\ket{100} + e^{\frac{2\pi i}{3}}\ket{010} + e^{\frac{4\pi i}{3}}\ket{001}}{\sqrt{3}}, \\ 
    \ket*{\tilde{W}^3_-} &= \frac{\ket{100} + e^{-\frac{2\pi i}{3}}\ket{010} + e^{-\frac{4\pi i}{3}}\ket{001}}{\sqrt{3}} 
    .
\end{align}
\end{subequations}
The three complex coefficients can be described with three real parameters after imposing the constraint that $\ket*{W^3}$ be a null state and by noting that we are always free to multiply jump operators by an overall phase without altering the resulting dissipation. We denote these parameters $\theta, \phi, \Gamma$, two angles and the square root of a rate, which are related to the dissipator coefficients via 
\begin{align}
    \pmqty{r_1[E] \\ r_2[E] \\ r_3[E]}
    =
    \frac{1}{\sqrt{3}}\pmqty{
    1 & 1 & 1 \\
    1 & e^{-\frac{2\pi i}{3}} & e^{ \frac{2\pi i}{3}} \\
    1 & e^{-\frac{4\pi i}{3}} & e^{ \frac{4\pi i}{3}} 
    }
    \pmqty{0 \\ \Gamma\cos\theta \\ e^{i\phi}\Gamma\sin\theta}
    .
    \label{eq:disscoeff}
\end{align}
In our simulations, we choose $\Gamma = 1$ so that the jump operator is normalized to $\sum_j |r_j[E]|^2 = 1$. We choose $\theta = 3\pi/4$ and $\phi = \pi/3$, leading to (after canceling an irrelevant phase)
\begin{align}
    r_1[E] = \frac{1}{\sqrt{6}},
    \qquad
    r_2[E] = \frac{1}{\sqrt{6}},
    \qquad
    r_3[E] = -\frac{2}{\sqrt{6}}
    .
    \label{eqn:CoeffsLMany}
\end{align}
We make this choice such that the jump operator coefficients are purely real, for simplicity. In Sec.~\ref{ssec:CompareConfigs}, we examine the impact the choice of coefficients can have for a large number of dissipator configurations.
%
%
%
%
\subsection{Hamiltonian}
%
We now turn our attention to the system Hamiltonian, $\HH_{\rm S}$. Through the remainder of this work we will examine three different forms of Hamiltonians: firstly, we look for the most resource-efficient Hamiltonian, i.e. the one which requires the fewest drives [minimum $n_{S}$ in Eq.~(\ref{eq:intro})].  We find that such a Hamiltonian has $2$ linear driving terms and $(2N-4)$ trilinear interactions, with
\begin{subequations}
\begin{alignat}{2}
    a_1 &= f_{2,1j} & &=  1 \\
    a_2 &= f_{1,2j} & &= -1 
    ,
\end{alignat}
    \label{eqn:CoeffsMin}%
\end{subequations}
and all other coefficients set to zero. This $\HH_{\rm S}$ is \emph{unique}, up to a permutation of the qubit indices or rescaling by an overall constant. We therefore refer to it as the ``minimal'' Hamiltonian. The linear scaling of resources, $n_{S} \sim \mathcal{O}(N)$, achieved by this minimal Hamiltonian is compatible with our goal of scalability as laid out in the introduction.
\par
For comparison, we also examine a ``maximal'' Hamiltonian which employs all $N$ possible linear drives and all $\binom{N}{3}$ possible generalized hopping interactions in Eq.~\eqref{eqn:FinalH}. The constraints placed on the coefficients are insufficient to uniquely specify them, so we choose them to be
\begin{align}
    a_j = f_{j,kl} = \exp\pqty{\frac{2\pi i }{N} j} 
    ,
    \label{eqn:CoeffsMax}
\end{align}
as a representative example. While careful optimization of these coefficients can yield better performance, we expect that the conclusions we draw about the scalability of our protocol and its performance should be generally insensitive to such fine tuning. 
\par
In addition to the minimal and maximal Hamiltonians defined above, we will also explore the performance of a Hamiltonian with $3$ linear drives and $(6N-9)$ generalized hopping interactions defined by the coefficients
\begin{subequations}
\begin{alignat}{5}
    a_1 &= f_{1,23} & &= f_{1,2j} & &= f_{1,3j} & &= 1 \\
    a_2 &= f_{2,13} & &= f_{2,1j} & &= f_{2,3j} & &= \exp\pqty{\frac{2\pi i}{3}} \\
    a_3 &= f_{3,12} & &= f_{3,1j} & &= f_{3,2j} & &= \exp\pqty{\frac{4\pi i}{3}} 
    ,
\end{alignat}
    \label{eqn:CoeffsNearMin}%
\end{subequations}
for $4\le j \le N$ and all other coefficients zero. We refer to this Hamiltonian as ``nearly-minimal''; notably, while it has the same linear resource scaling as the minimal Hamiltonian, we will show that the performance of the protocol with this Hamiltonian is similar to that obtained with the maximal Hamiltonian, indicating that efficient resource scaling and good performance can coexist in our design.
%
\section{Numerical results}
\label{sec:Results}
%
An important performance metric of interest for a stabilization protocol is the rate ($\tau^{-1}$) at which it prepares the target state. As discussed earlier in Sec.~\ref{ssec:TargetState} for stabilization protocols with zero intrinsic error, the steady state fidelity is primarily set by the stabilization rate in the presence of realistic decoherence. The scaling of the stabilization rate with the number of qubits provides a good measure of the scalability of the protocol, as it must remain sufficiently high as to remain in the dissipation engineering regime ($\tau^{-1}\gg\gamma^W_\downarrow$) for stabilization to be effective. 
\par
For the results presented in this work, we numerically simulate the master equation,
\begin{align}
    \LV_{\rm eng}\dop(t) = 
    \dv{\dop(t)}{t} = 
    -i\comm{\lambda\HH^{(3)}}{\dop(t)} 
    + \sum_{j=1}^m \mathcal{D}\left[\LL_{E_j}\right] \dop(t)
    ,
    \label{eqn:SchemeLV}
\end{align}
where $\{E_1,\dots,E_m\}$ denote the $m$ hyperedges describing the dissipator configuration and $\lambda$ is an arbitrary scale factor weighting the Hamiltonian and dissipative couplings in our protocol. Without loss of generality, we set $\lambda = 1/4$ in all simulations in order to have the engineered dissipation be the dominant component of our protocol.
\par
The instantaneous error (infidelity) at time $t$
\begin{align}
    \varepsilon(t) = 1 - \left|\mel*{W^N}{\dop(t)}{W^N}\right|^2
    ,
\end{align}
decays exponentially (ignoring the initial-state dependent transient behavior at short times) as
\begin{align}
    \varepsilon(t) \approx \varepsilon_0 \exp\pqty{-t/\tau}
    ,
\end{align}
where the stabilization time constant $\tau$ characterizes the rate at which stabilization occurs. 
\par
Our simulations employ a combination of the Quantum Toolbox in Python (QuTiP) framework \cite{QuTiP} for few-qubit simulations and a custom C++ solver driven by QuTiP code for larger simulations. The output of each simulation is a trace of $\varepsilon(t)$ from $t=0$ to some large $t_{\rm max}$ which is then fit to extract a numerical estimate of $\tau$. Additionally, in the presence of local qubit decoherence, we extract the steady-state error $\varepsilon_\infty \approx \varepsilon(t_{\rm max})$ to verify that the stabilization process does in fact converge to the $\ket{W^N}$ as expected.
%
\subsection{Performance scaling with system size}
\label{ssec:Scaling}
%
%
\begin{figure*}[t!]
    \centering
    \includegraphics[width=\textwidth]{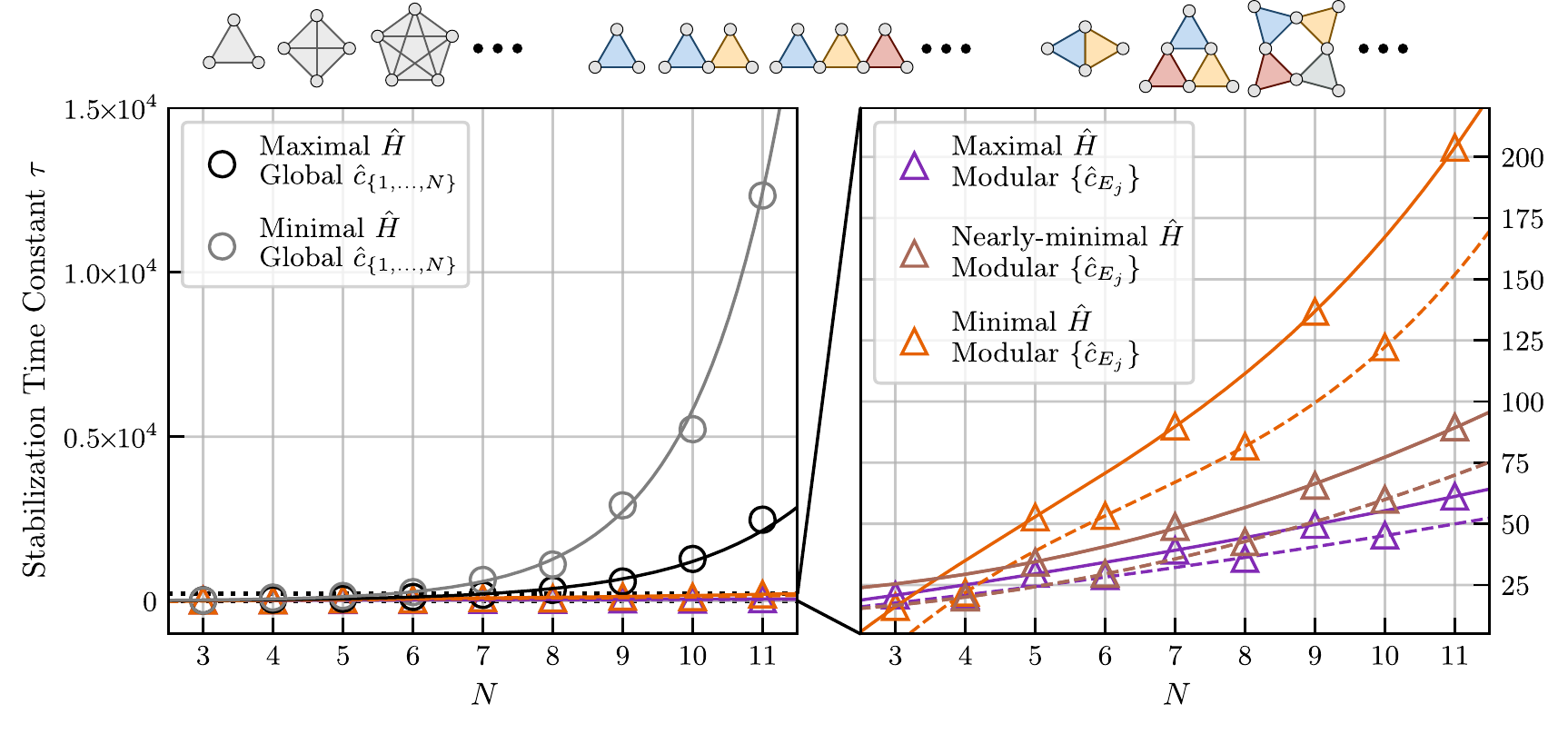}
    \caption{Scaling of the stabilization time constants with the number of qubits $N$, obtained from simulations with various Hamiltonians and dissipator configurations. The dissipator hypergraphs are depicted above the plots, from left to right they are: a single global width-$N$ dissipator in gray, followed by the modular chain and ring configurations in color. (Left) The simulation results obtained for the single global dissipator with both the maximal and minimal $\HH_{\rm S}$ are shown in black and gray, with the solid lines showing exponential fits of $\tau(N)$. (Right) Zoom to show the scaling with overlapping dissipator configurations. In the modular case, $\tau(N)$ is well-fit by a polynomial, quadratic when using the maximal and nearly-minimal $\HH_{\rm S}$ and cubic for the minimal $\HH_{\rm S}$. Fits for odd and even $N$ are shown as solid and dashed lines, respectively.}
    \label{fig:scaling}
\end{figure*}
Figure~\ref{fig:scaling} shows the stabilization time constant $\tau$ extracted from simulations across $N=3$ to $N=11$ qubits for five versions of our protocol. The two panels of this figure show the results for the two distinct classes of engineered dissipators, namely:
\begin{itemize}
    \item A global width-$N$ dissipator [Eq.~\eqref{eqn:CoeffsLOne}], combined with either the maximal [Eq.~\eqref{eqn:CoeffsMax}] or minimal [Eq.~\eqref{eqn:CoeffsMin}] $\HH_{\rm S}$.
    \item Minimal connected configurations of width-3 dissipators in the chain/ring configurations for odd/even $N$ [Eq.~\eqref{eqn:CoeffsLMany}], combined with the maximal, minimal, or nearly-minimal [Eq.~\eqref{eqn:CoeffsNearMin}] $\HH_{\rm S}$.
\end{itemize}
\par
In the first case with a single global dissipator, we find that $\tau$ appears to grow exponentially with the number of qubits. In contrast, with modular overlapping dissipators, $\tau$ increases only polynomially with the number of qubits -- a significant improvement over the scaling exhibited with a single global dissipator. Notably, this difference in scaling persists with different types of $\HH_{\rm S}$: specifically, the scaling of $\tau$ with $N$ is well-fit by a cubic polynomial when using the minimal Hamiltonian, improving to a quadratic polynomial in $N$ when using the maximal Hamiltonian. The nearly-minimal Hamiltonian also exhibits quadratic scaling of $\tau$ with $N$, illustrating that the majority of the performance benefit associated with the maximal Hamiltonian can be reclaimed with a modest increase in the number of drives while maintaining linear scaling of resources in system size.
%
%
%
\par
To understand the stark difference in the scaling of $\tau$ between the two approaches to dissipator construction, it is instructive to consider what changes as the number of qubits $N$ is increased. For a single global dissipator, both the Hilbert space on which it acts and the number of stationary states grow rapidly with $N$. On the other hand, for the modular dissipator construction, the $k$-qubit Hilbert space on which each dissipator acts remains the same regardless of $N$. There may be additional global effects arising from interactions between overlapping dissipators. However, as $N$ grows we expect such effects to become less relevant for the ring and chain configurations studied here as these configurations minimize the number of qubits shared between dissipators. 
\par
Though the results we have shown for the modular dissipator construction only employ width-$3$ dissipators, we observe qualitatively similar results for other dissipator widths as well. Appendix~\ref{app:Width4} shows the scaling of the stabilization time with $N$ for modular configurations with width-$4$ dissipators, which we find to be significantly closer to that of the width-$3$ modular configurations shown in Fig.~\ref{fig:scaling} than the exponential scaling of the global configuration. We expect this to be similar for any fixed width, as the relevant distinction is between modular (fixed-width) and global (increasing-width) dissipation.
%
\subsection{Modular construction with \texorpdfstring{$N=5, k=3$}{N=5, k=3}}
\label{ssec:CompareConfigs}
%
\begin{figure*}[t!]
    \centering
    \includegraphics[width=\textwidth]{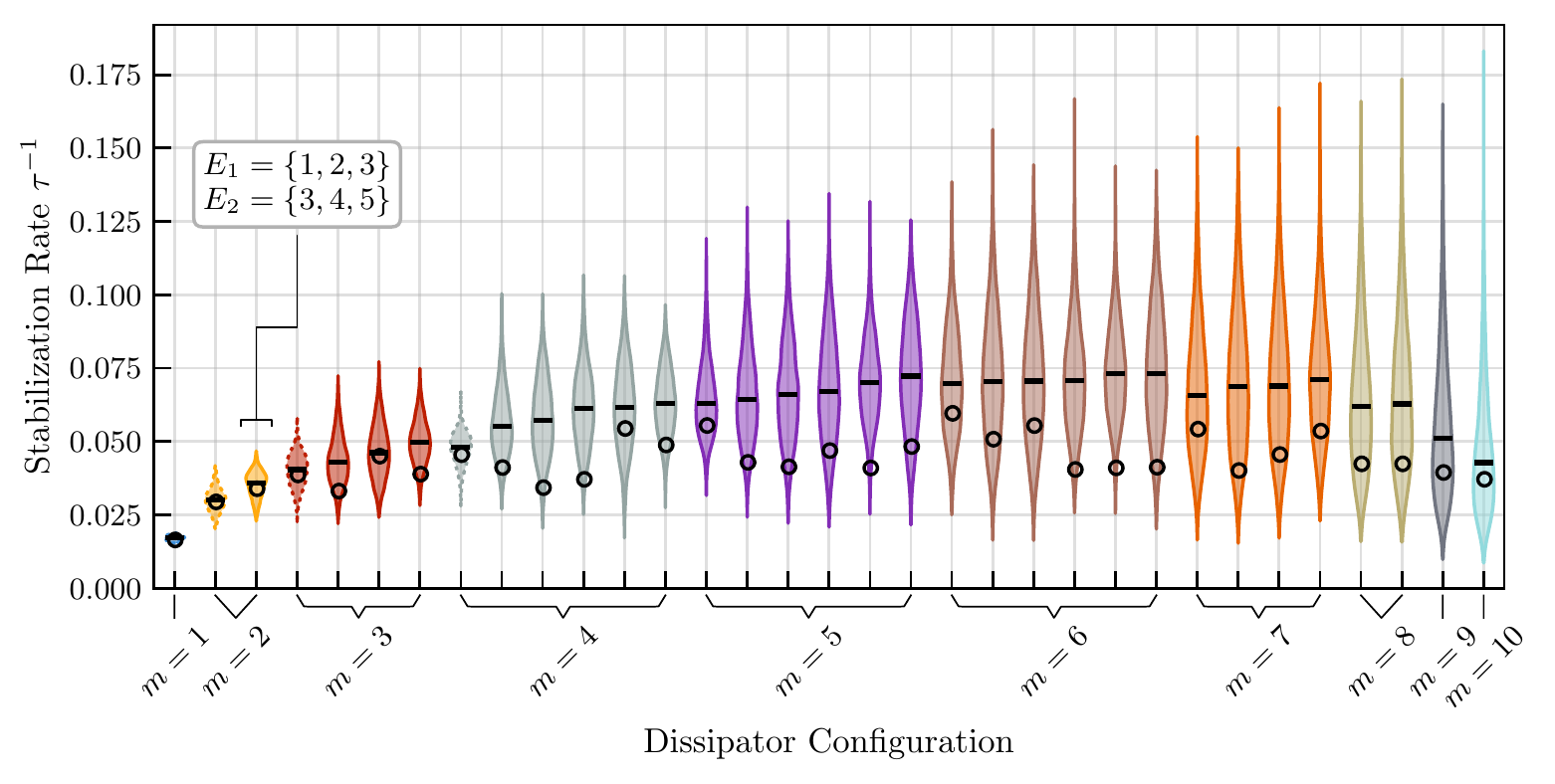}
    \caption{Stabilization rates for all 33 distinct configurations of width-$3$ dissipators for $N=5$, where $m$ denotes the number of dissipators. The sets $\{E_j\}$ defining each configuration are listed in Table~\ref{tab:Configs} in Appendix~\ref{app:EnumerateFive}, in the order they appear in this figure. Each configuration was simulated with all jump operators independently randomized 10240 times. The distributions of stabilization rates are plotted vertically, with solid outlines for connected configurations and dotted outlines for disconnected configurations. The median of each distribution is marked with a black bar, and the circles indicate the performance achieved using the same $(\theta,\phi) = (3\pi/4,\pi/3)$ for each dissipator, as in Fig.~\ref{fig:scaling}.}
    \label{fig:5qbConfigs}
\end{figure*}
In the previous section, we examined how $\tau$ varies with the number of qubits $N$ for a few fixed families of dissipator and Hamiltonian configurations. For the modular dissipator construction, we considered only the chain and ring configurations. To study the impact of this choice, we consider all the different possible protocols for stabilizing $\ket{W^5}$ with width-$3$ dissipators and the maximal Hamiltonian. On five nodes (qubits), there are $\binom{5}{3} = 10$ possible width-$3$ hyperedges (dissipators), which may be combined into 33 inequivalent hypergraphs (dissipator configurations). As the maximal Hamiltonian we employ is symmetric with respect to permutations of the qubits, we expect that different dissipator configurations which can be related by re-labelling the qubits will lead to the same performance. Hence, we only need to consider configurations corresponding to distinct hypergraphs. For details, see Appendix~\ref{app:EnumerateFive}.
\par
Having chosen a configuration described by a set of jump operators $\{\LL_{E_j}\}$, we still need to choose the coefficients $r_k[E_j]$. For the simulations shown previously in Sec.~\ref{ssec:Scaling}, we chose to use the same coefficients for each dissipator, corresponding to $(\theta_j,\phi_j)=(3\pi/4,\pi/3)$. In a practical implementation, we expect that each dissipator would be independently tuned and so the corresponding parameters will likely vary. We can model this by independently choosing the $(\theta_j,\phi_j)$ pair for each jump operator randomly from a uniform distribution. Figure~\ref{fig:5qbConfigs} shows the distribution of stabilization rates obtained for each of the 33 distinct dissipator configurations, where each configuration was sampled over 10240 random choices of the parameters. 
\par
The results of these simulations are noteworthy in several respects. Firstly, and most importantly, we found that the protocol successfully stabilized the target state $\ket{W^5}$ for every set of parameters tested. This indicates that the stabilization mechanism itself is robust and insensitive to the specific parameter choices made for each dissipator. Additionally, this confirms our earlier statement that it is exceedingly unlikely that any eigenstates of the Hamiltonian are stationary states of the dissipation, other than the target state. Secondly, we find that the distributions of stabilization rates are fairly tightly centered around the median (indicated with a black bar) when the number of dissipators $m$ is small. For these more resource-efficient configurations, the performance is primarily set by the structure of the hypergraph rather than the detailed parameter choice. In particular, we see that the choice of parameters used in the scaling simulations (indicated with a black circle) are representative. In contrast, for configurations with many dissipators we see that the distributions of stabilization rates are no longer sharp and instead exhibit long tails, generally towards higher performance. From this, it is clear that optimization of each dissipator will be beneficial when employing many dissipators with significant overlap.
\par
An interesting feature of the simulation results presented in Fig.~\ref{fig:5qbConfigs} is that the median stabilization rate does not increase monotonically with the number of dissipators $m$. Intuitively, it might have been expected that the addition of an extra dissipator would always lead to faster stabilization, however this is not the case. We suspect that this due to the increasing overlap between the various dissipators with increasing $m$. The actions of two overlapping dissipators generally will not commute, and can generate additional non-trivial dynamics in the relaxation process to the stationary subspace, leading to slower stabilization. For hypergraphs with significant overlap (e.g., the maximal hypergraph with $m=10$), this effect is most pronounced and outweighs the benefit of the additional dissipators. The long tails towards higher stabilization rates exhibited by such configurations indicates that optimization can mitigate this issue, however the observed sub-linear best-case growth of $\tau^{-1}(m)$ appears to indicate that these effects cannot be entirely avoided. 
\par
Note that there are three disconnected configurations shown in Fig.~\ref{fig:5qbConfigs}, one each with $m=1,2,3$. These configurations have large and poorly-constrained stationary spaces. Therefore, when using one of these configurations the protocol relies heavily on the Hamiltonian to exclude these states and select $\ket*{W^N}$ as the unique target state. The comparatively narrow distributions of stabilization rates obtained for a disconnected dissipator configuration indicate that the details of the dissipation are less important in this case. An extreme example is provided with the $m=1$ configuration consisting of a single width-$3$ dissipator. In this case, two qubits see no dissipation at all and so their state is entirely unconstrained by the dissipation. However, the maximal Hamiltonian employed in these simulations is a complex and highly nonlinear operator, with little degeneracy and complicated eigenstates that span most of the Hilbert space. Thus, we find that even in this case, it is very unlikely that any extra eigenstates of the maximal Hamiltonian are stationary under the single dissipator. We do \emph{not} not find that stabilization is possible in these disconnected cases with either the minimal or nearly-minimal Hamiltonians; their eigenstates have a simpler structure, generally leading to several degenerate eigenstates being stationary states of the dissipation. 
\par
More generally, connected configurations of dissipators of any width will allow stabilization of $\ket*{W^N}$ with any of the three Hamiltonians we have defined, while disconnected configurations will only work with the maximal Hamiltonian. For instance, in Appendix~\ref{app:Width4}, we repeat this analysis using width-$4$ dissipators to stabilize $\ket*{W^5}$. Again, with the maximal Hamiltonian we find that every randomized simulation for each dissipator configuration was successful in stabilizing $\ket*{W^5}$. As in the width-$3$ case, we expect that disconnected configurations will not work with the more resource-efficient minimal and nearly-minimal Hamiltonians. 
%
\subsection{Robustness to decoherence}
\label{ssec:Decoherence}
%
\begin{figure}[t!]
    \centering
    \includegraphics[width=\columnwidth]{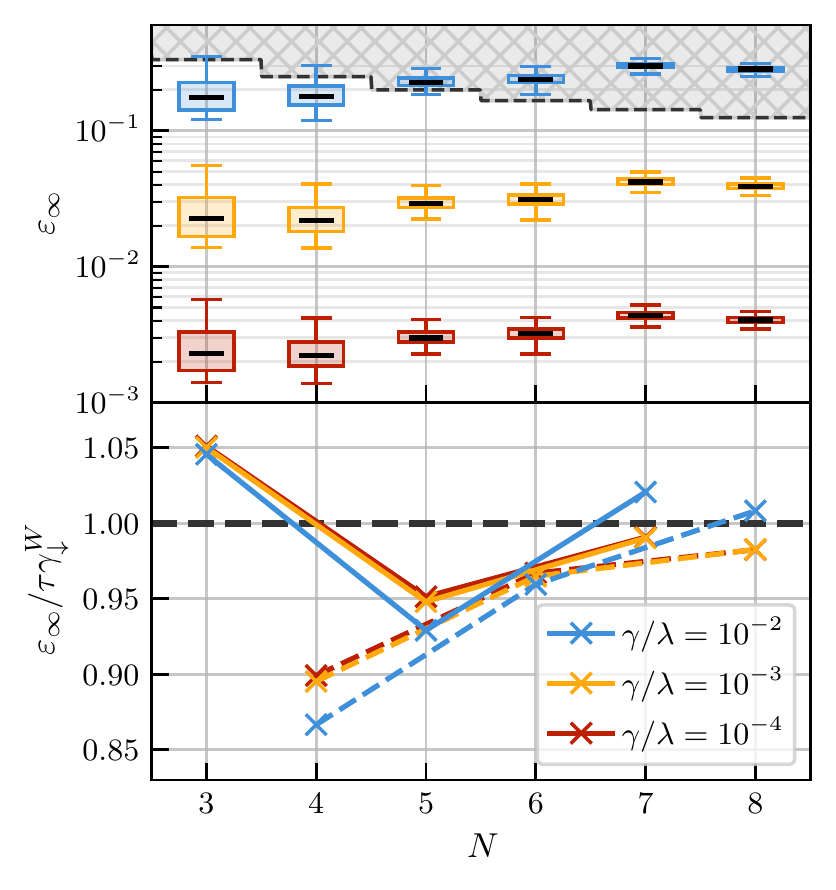}
    \caption{(Top) Box-and-whisker plots showing the distributions of achievable steady-state error, $\varepsilon_{\infty}$, (with the median highlighted in black), simulated with the maximal $\HH_{\rm S}$ and the chain (odd $N$) or ring (even $N$) dissipator configurations, using randomized coefficients and three different choices of $\gamma^z = \gamma^- \equiv \gamma$. In the gray region, $\varepsilon_{\infty}$ is too high for the projector witness \eqref{eqn:ProjectorWitness} to certify the presence of $N$-qubit entanglement. (Bottom) Plot of the ratio of the simulated $\varepsilon_\infty$ to the estimate from our simple rate model, $\tau\gamma^{W}_\downarrow$, with the dissipator parameters in Eq.~\eqref{eqn:CoeffsLMany}.}
    \label{fig:decoherence}
\end{figure}
As discussed in Sec.~\ref{ssec:TargetState}, stabilization of $\ket{W^N}$ with unit fidelity is not possible in the presence of uncontrolled decoherence. Nonetheless, provided the decoherence remains `weak' compared to the engineered dissipation, the protocol can still converge to a unique steady state $\dop(t\to\infty)$ which, while not pure, can have a very high overlap with the target state. In this section, we study the effect of decoherence on our protocol, described by local relaxation and dephasing on each qubit. Specifically, we consider the master equation,
\begin{align}
    \dv{\dop(t)}{t} = \mathcal{L}_{\rm eng}\dop(t)
    + \sum_{j=1}^N\Big(
        \gamma^{z}_j\mathcal{D}\left[\sigz_j\right]
        +
        \gamma^{-}_j\mathcal{D}\left[\sig_j\right]
    \Big)\dop(t)
    ,
    \label{eqn:SchemeDecayLV}
\end{align}
where $\mathcal{L}_{\rm eng}\dop(t)$ describes the engineered dynamics given by Eq.~\eqref{eqn:SchemeLV}, and $\gamma^{-}_j, \gamma^{z}_j$ are the local relaxation and dephasing rates, respectively. As discussed in Appendix~\ref{app:WDecay}, in the dissipation engineering regime the dynamics can be approximated as a competition between feeding into $\ket*{W^N}$ at a rate $\tau^{-1}$ and decay out of it at a rate $\gamma^W_\downarrow$. The approximate steady-state error is then $\varepsilon_\infty \approx \tau\gamma^W_\downarrow$, which we expect to scale linearly with the decoherence rates.
\par
Figure~\ref{fig:decoherence} shows the results of simulations of Eq.~\eqref{eqn:SchemeDecayLV}, with $\mathcal{L}_{\rm eng}$ including the maximal Hamiltonian and the same chain and ring dissipator configurations employed in Sec.~\ref{sec:Results}a. In the top panel, each box-and-whisker plot shows the distribution of steady-state errors obtained from 10240 simulations with randomized parameter choices for each dissipator, as described in Sec.~\ref{ssec:CompareConfigs}. We performed simulations for three different choices of decoherence rates, where in each case all the local decoherence rates were set equal to one another, $\gamma^{-}_j = \gamma^{z}_j \equiv \gamma$. 
The three cases we have simulated range from pessimistic, with $\gamma = \lambda\times10^{-2}$, to optimistic with $\gamma = \lambda\times10^{-4}$, where $\lambda$ controls the strength of the engineered Hamiltonian in Eq.~\eqref{eqn:SchemeLV}. 
Note that the $\varepsilon_{\infty}$ remains the same for different local decoherence rates, as long as they lead to the same effective $\gamma^W_\downarrow$ (Appendix~\ref{app:GammaDist}).
\par
Note that the distributions of steady-state errors become sharper as $N$ increases. With the chain and ring dissipator configurations employed in these simulations, we expect that as the number of qubits grows, the details of each individual dissipator matter less to the overall stabilization rate (and hence, steady-state error).
\par
It is possible to certify the existence of genuine $N$-qubit W-like entanglement in the steady state using the projector witness \cite{Toth2009} given by
\begin{align}
    \hat{\mathcal{W}}_W = \left(\frac{N-1}{N} \right)\mathbbm{1} - \dyad*{W^N}{W^N}
    ,
    \label{eqn:ProjectorWitness}
\end{align}
whose expectation value $\expval*{\hat{V}_W}_\infty = \Tr\!\left.[\hat{\mathcal{W}}_W\dop(t\to\infty)]\right.$ is a function solely of the steady-state error
\begin{align}
    \expval{\hat{\mathcal{W}}_W}_\infty = \varepsilon_\infty - \frac{1}{N}
\end{align}
If the steady state is such that $\expval*{\hat{\mathcal{W}}_W}_\infty < 0$, then it is sufficiently close to $\ket{W^N}$ to be guaranteed to possess $N$-qubit entanglement. While more sophisticated entanglement witnesses and verification methodologies could provide more refined estimates of the amount of usable entanglement present, the projector witness in Eq.~(\ref{eqn:ProjectorWitness}) suffices to provide a straightforward and unambiguous threshold for understanding the limits of the scalability of our protocol in the presence of realistic qubit decoherence.
\par
In the gray region shown in the top panel of Fig.~\ref{fig:decoherence}, $\varepsilon_\infty$ is too high for the projector witness to certify the existence of $N$-qubit entanglement. For most parameter choices, we find that in the pessimistic case where $\gamma/\lambda = 10^{-2}$, our protocol is able to reliably generate W-like entanglement for 3 or 4 qubits. For a more reasonable choice of $\gamma/\lambda = 10^{-3}$ (or lower), simulations show that even with randomly chosen parameters our protocol yields errors sufficiently low to certify the existence of $N$-qubit entanglement out to $N=8$ qubits, and that it will do so for somewhat larger $N$ as well. We conclude that local decoherence will thus not be a serious impediment to the implementation of our protocol.
\par
As expected from our rate model estimate, we see that the minimum achievable $\varepsilon_\infty$ decreases $\sim10\times$ when the decoherence rates are reduced by the same factor. The bottom panel of Fig.~\ref{fig:decoherence} directly compares $\varepsilon_{\infty}$ to the simplistic estimate from the rate model discussed in Appendix~\ref{app:WDecay}, $\epsilon_{\infty}\approx \tau\gamma^W_\downarrow$. We find this estimate to be accurate to within $\sim15\%$, improving as the number of qubits grows. This justifies our focus on the stabilization rate as the primary performance metric, as the steady-state error can be simply estimated from it once a decoherence model is specified.
%
%
\section{Discussion and Conclusions}
\label{ssec:Conclusion}
%
In summary, in this work we have given an operational definition of ``scalability'' for dissipation engineering protocols and, using it as a guiding principle, shown how scalable entanglement stabilization of W-states is accessible using linear dissipation and trilinear Hamiltonians. Crucially, we find that employing a modular approach to dissipation engineering with many overlapping few-qubit dissipators leads to polynomial scaling of the stabilization time, whereas employing a single global dissipator leads to exponential growth in stabilization time. Interestingly, our results offer a counterexample, albeit from a dissipation engineering perspective, to the conventional wisdom of maximizing coupling connectivity for improving efficacy of control in multi-qubit systems. 
\par
In addition to superior performance, the modular dissipation approach proposed here can offer several practical advantages from the point of view of experimental implementations. First, we expect the additional hardware cost required to have several auxiliary modes as engineered baths to be offset by the additional simplification accompanying modular design and fabrication, along with enhanced flexibility in engineering qubit interactions with each width-$k$ dissipator independently. Further, restricting to linear engineered dissipation while designing our protocols also makes them amenable to current experimental techniques in several platforms such as superconducting circuits \cite{Gu2017}, trapped ions \cite{Leibfried2003} and atoms \cite{Kaufman2012}, and optomechanical systems \cite{Aspelmeyer2014}. Specifically, parametrically-coupled circuit-QED systems provide a promising direction for implementation of our protocol, since significant control over the form, amplitudes, and phases of the various interaction terms in the Hamiltonian will be crucial to implementing the desired system Hamiltonian \cite{Hillman2020,Ye2021,Menke2022}. A specific example can be found in our recent work \cite{Brown21}, where dissipative stabilization of a Bell state was performed in a parametric circuit-QED system with engineered linear dissipation realized via pumping qubit-resonator sidebands. 
\par
Another practical advantage of our proposed stabilization protocol is its robustness to undesired local decoherence, as exemplified by results presented in Fig.~\ref{fig:decoherence}. A significant contributory factor here is the robustness of $\ket*{W^N}$ to local decoherence especially for large $N$; this advocates taking into account the properties of the target quantum state very early on in the design process while considering them as candidates for dissipative state preparation. In conjunction with the favorable polynomial scaling of stabilization time with modular dissipation engineering, this can inform design of state preparation protocols in large noisy quantum systems.
\par
The modular approach to dissipation engineering proposed here also suggests a number of possible extensions of this work. In particular, it would be interesting to consider a more general class of engineered jump operators, for example those describing nonlinear or conditional dissipation. Such jump operators could allow the stabilization of a broader class of entangled states, such as GHZ or cluster states, or potentially allow even more resource-efficient stabilization of the $\ket*{W^N}$ states considered here. One could also consider expanding the system from a collection of qubits to a collection of qudits, either for the purposes of stabilizing $N$-qudit states or as a means for introducing auxiliary states for dissipation engineering \cite{Brown21,Sorensen16}. 
\par
An additional possible extension of this work would be to employ ``parallelization and quantum feedback'' where, instead of stabilizing an $N$-qubit state directly, several copies of the stabilization protocol are run on small subsets of qubits and the resulting entangled states are ``fused'' together using few-qubit measurements and quantum feedback. This approach was presented in \cite{Lukin2021} and yielded an exponential speedup in the preparation time (polynomial to logarithmic) required by a protocol which stabilizes AKLT states.  We expect that similar speedups could be achievable with our protocol, further improving the achievable performance.
\par
In any of these scenarios, once a target state has been specified and a choice has been made as to the generic form of the engineered dissipation and interactions entering the Hamiltonian, the same constraint-driven design process undertaken in this work can be used to determine if stabilization is possible, and if so how complex the resulting protocol must be. Recently, a similar method was employed for numerical discovery of ``$\sqrt{3}$" bosonic code that corrects for excitation loss using only low-depth Hamiltonians \cite{Wang2022}. Deploying such computational techniques for optimizing and identifying minimal architectures for modular dissipation engineering of large quantum systems is a clear direction for future works to explore.
%
\section*{Acknowledgments}
%
This research was supported by the Air Force Office of Scientific Research under grant FA9550-21-1-0151 and Department of Energy under grant DE-SC0019461. E.D. acknowledges support from the Quantum Information Science and Engineering Network (QISE-NET) Fellowship funded by NSF award DMR-1747426. 
Some of the simulations presented in this work were performed with resources provided by the Massachusetts Green High Performance Computing Cluster (MGHPCC). The color scheme we use for our figures is taken from \cite{Petroff21}. 
\appendix
%
\section{W-state decay rate}
\label{app:WDecay}
%
We consider a system of $N$ qubits whose dynamics are described by the Lindbladian
\begin{align}
    \LV
    &\equiv 
        -i\comm{\HH}{\sdot}
        + \sum_{k=1}^m \Gamma_k \mathcal{D}\left[\LL_{E_k}\right]\sdot \nonumber\\
        &\quad
        + \sum_{j=1}^N \gamma_j^{-} \mathcal{D}\left[\sig_j\right]\sdot
        + \sum_{j=1}^N \gamma_j^{z} \mathcal{D}\left[\sigz_j\right]\sdot \nonumber\\
    &\equiv
        \LV_\uparrow+ \LV_\downarrow
    .
\end{align}
The first two terms, collected into $\LV_\uparrow$, describe the stabilization process which attempts to drive the system to some pure steady state $\kpsi$. The remaining two terms, collected into $\LV_\downarrow$, describe local amplitude damping and dephasing processes which occur on each qubit. Assuming that the system operates in the dissipation engineering regime, where the stabilization rate is much stronger than the decoherence processes, the long time steady state of the system is very close to $\kpsi$ with an error approximately set by the competition of the feeding and decay rates, $\varepsilon_\infty\approx\gamma_\downarrow/\Gamma_\uparrow$.
\par
As the system is nearly in the steady state of $\LV_\uparrow$, the rate relevant to the error estimate $\Gamma_\uparrow \equiv \tau^{-1}$ is the rate governing the late-time exponential approach to the steady state, or the inverse of the stabilization time constant. This is approximately the real part of the spectral gap of $\LV_\uparrow$. Conversely, at late times, the system will be very far from the steady state of $\LV_\downarrow$. The effective decay rate $\gamma_{\downarrow}$ should be identified with the decay rate of $\kpsi$ under the action of $\LV_\downarrow$ at short times, described by the propagator $V_\downarrow(t_f,t_i)\equiv\exp[(t_f-t_i)\LV_\downarrow]$. We can find the effective decay rate by calculating the overlap between the target state $\hat{\Pi}_\psi$ with the same state propagated forward for a short time interval $\epsilon$ under the influence of the decoherence,
\begin{align}
      \tr\left[\hat{\Pi}_\psi V_\downarrow(\epsilon,0)\hat{\Pi}_\psi\right] 
    &= \tr\left[\hat{\Pi}_\psi e^{\epsilon \LV_\downarrow}\hat{\Pi}_\psi\right] \nonumber\\ 
    &\approx \tr\left[\hat{\Pi}_\psi \left(1 + \epsilon\LV_\downarrow\right)\hat{\Pi}_\psi\right] \nonumber\\ 
    &= \tr\left[\hat{\Pi}_\psi \left(\hat{\Pi}_\psi + \epsilon\LV_\downarrow\hat{\Pi}_\psi\right)\right] \nonumber\\ 
    &= 1 + \epsilon\tr\left[\hat{\Pi}_\psi\LV_\downarrow\hat{\Pi}_\psi\right] \nonumber\\
    &= 1 - \epsilon \gamma_{\downarrow}
    ,
\end{align}
where we have identified $\gamma_\downarrow \approx -\tr\small[\hat{\Pi}_\psi\LV_\downarrow\hat{\Pi}_\psi\small]$. Note that since $\LV_\downarrow$ is a superoperator, we cannot use the cyclic property of the trace to directly simplify this further.
Substitution of the definition of $\LV_\downarrow$ and of $\hat{\Pi}_\psi = \dyad{\psi}{\psi}$ yields
\begin{align}
    \gamma_\downarrow 
    &\approx -\tr\Big[\dyad{\psi}{\psi}\LV_\downarrow\dyad{\psi}{\psi}\Big] \nonumber\\
    &= -\bra{\psi}\Big(\LV_\downarrow\dyad{\psi}{\psi}\Big)\ket{\psi} \nonumber\\
    &= -\sum_{j=1}^N \gamma_j^{-} \mel{\psi}{\Big(\mathcal{D}\left[\sig_j\right]\dyad{\psi}\Big)}{\psi} \nonumber\\
        &\quad
        - \sum_{j=1}^N \gamma_j^{z} \mel{\psi}{\Big(\mathcal{D}\left[\sigz_j\right]\dyad{\psi}\Big)}{\psi}.
    \label{eqn:GammaDown}
\end{align}
To ease our computation further we only consider states $\ket{\psi}$ that are symmetric under arbitrary qubit permutations, as then each sum is simply $N$ times its summand. We write $\ket{\psi}$ in the following form which distinguishes the first qubit,
\begin{align}
    \ket{\psi} &= \alpha\ket{0}\ket{\phi_0} + \beta\ket{1}\ket{\phi_1} 
    ,
    \label{eqn:AsymPsi}
\end{align}
where $\ket{\phi_0}$ and $\ket{\phi_1}$ are arbitrary states on the remaining $N-1$ qubits which need not be orthogonal. Both $N$-qubit GHZ and W states have simple expressions in this form
\begin{subequations}
\begin{align}
    \ket{W^N} &= \sqrt{\frac{N-1}{N}}\ket*{0}\!\!\ket*{W^{N-1}} + \frac{1}{\sqrt{N}}\ket*{1}\!\!\ket*{0^{N-1}}, \\
    \ket{{\rm GHZ}^N} &= \frac{1}{\sqrt{2}}\ket*{0}\!\!\ket*{0^{N-1}} + \frac{1}{\sqrt{2}}\ket*{1}\!\!\ket*{1^{N-1}}.
\end{align}
\end{subequations}
Substitution of \eqref{eqn:AsymPsi} into \eqref{eqn:GammaDown} and some simplification yields a generic expression for the decay rate of a permutation symmetric state under $\LV_\downarrow$,
\begin{align}
    \gamma_\downarrow &\approx 
    |\beta|^2 \Big(1 - |\alpha|^2 |\ip{\phi_1}{\phi_0}|^2\Big)\sum_{j=1}^N \gamma_j^{-} \nonumber\\
    &\quad
    + 4\Re{\left(\alpha\beta^*\right)^2}\sum_{j=1}^N \gamma_j^{z}
    .
    \label{eqn:GenericDecayRate}
\end{align}
If we take all the decay rates to be equal ($\gamma^-_j \to \gamma^-$, $\gamma^z_j \to \gamma^z$) and evaluate this decay rate for the $N$-qubit W and GHZ states, we find
\begin{subequations}
\begin{align}
    \gamma_\downarrow^{W} &\approx \gamma^{-} + 4\frac{N-1}{N}\gamma^{z}, \\ 
    \gamma_\downarrow^{\rm GHZ} &\approx \frac{N}{2}\left(\gamma^- + 2\gamma^z\right).
\end{align}
\end{subequations}
Note that the decay rate of $\ket*{W^N}$ is bounded as ${N\to\infty}$. 
\begin{figure}[t!]
    \centering
    \includegraphics[width=\columnwidth]{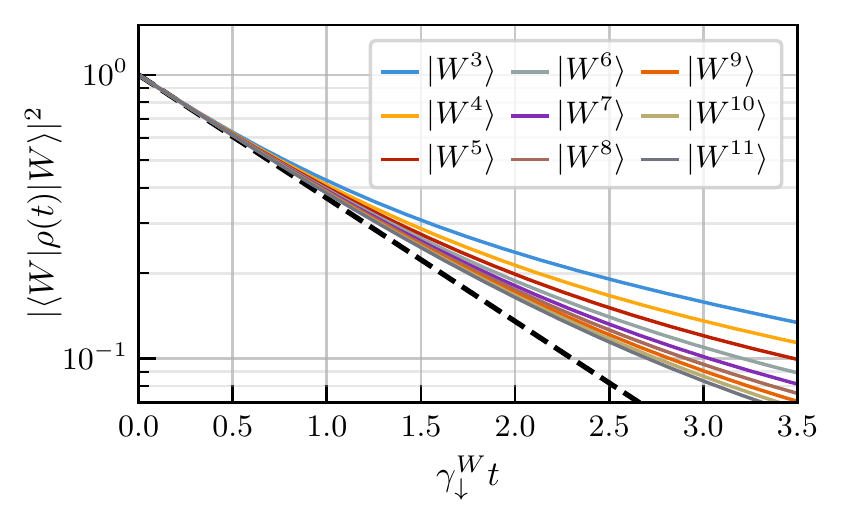}
    \caption{Master equation simulations of the decay of $\ket{W^N}$, for $3\le N\le11$ subject to local decoherence described by $\LV_{\downarrow}$, with all $\gamma^-_j = \gamma^z_j = \textrm{const}$. The black dashed line what we would expect to see if the $\ket{W^N}$ decayed purely exponentially at $\gamma^{W}_\downarrow(N)$. At short times this agrees well with the simulations as expected, while at later times the simulations show that the W state decays less quickly than our estimate would predict.}
    \label{fig:wdecay}
    \vspace{-10pt}
\end{figure}
Intuitively, this is a result of two features of the state $\ket*{W^N}$. Firstly, it is a single-excitation state for any value of $N$. That single excitation is delocalized among all the constituent qubits and therefore decays according to the average of the local damping rates. By contrast, the $\ket{{\rm GHZ}^N}$ has an average $N/2$ excitations which is reflected in its decay rate. Secondly, the act of flipping the phase of a single qubit in a system intialized in $\ket{W^N}$ does not produce an orthogonal state, and in fact as $N\to\infty$ the perturbed state becomes less and less distinguishable from the initial state. Again this is in stark contrast to the $\ket{{\rm GHZ}^N}$, where a phase-flip on any individual qubit produces an orthogonal state.
%
%
\section{Overlapping dissipator construction}
\label{app:OverlappingDissipators}
%
In this appendix, we detail how several overlapping few-qubit dissipators jointly select $\ket*{W^N}$ and $\ket*{0^{\otimes N}}$ as dark states. For simplicity we choose to consider the case of width-3 dissipators, however the logic generalizes easily to any combination of widths.
\par
As discussed in the main text, writing $\ket*{W^N}$ in the asymmetric form
\begin{align}
    \ket{W^N} &= \sqrt{\frac{N-3}{N}}\ket{000}\ket*{W^{N-3}} \nonumber\\
        &\quad
        + \sqrt{\frac{3}{N}}\ket*{W^3}\ket*{0^{\otimes N-3}}
    ,
\end{align}
allows us to easily see that a three-qubit jump operator $\LL_E$ acting on the qubits in $E=\cbqty{j,k,l}$ will support $\ket*{W^N}$ as a dark state if
\begin{align}
    \LL_E\ket{000} = \LL_E\ket*{W^3} = 0 
    .
\end{align}
In the main text we discuss one approach towards implementing such jump operators with linear dissipation, however in this appendix we choose to ignore the detailed implementation of $\LL_E$ and instead consider it to be some operator which has $\ker(\LL_E)=\mathrm{span}\cbqty{\ket{000},\ket*{W^3}}$ exactly. The jump operators engineered in the main text may support additional stationary states individually, however we find that these additional states do not cause any issues and only serve to complicate the discussion.
\par
We can show that if we have a configuration of dissipators that jointly constrain the space of stationary states to $\mathrm{span}\cbqty{\ket*{0^{\otimes N-2}}, \ket*{W^{N-2}}}$ (e.g., by assumption we have that $\LL_{\cbqty{1,2,3}}$ selects a $3$-qubit W state, so we may always begin with just a single dissipator), then by the addition of one more width-3 dissipator we are left with a configuration of dissipators on $N$ qubits whose kernel is again spanned by only $\ket*{W^N}$ and $\ket*{0^{\otimes N}}$.
\par
We show this by assuming that we have some initial configuration of dissipators which only support states of the form
\begin{align}
    \ket{\xi} = \alpha\ket*{W^{N-2}} + \beta\ket*{0^{\otimes N-2}}
\end{align}
as stationary states. Starting with such an arbitrary combination, we then expand the Hilbert space by adding two qubits. These two qubits could be entangled with the existing qubits, meaning the most general state including the two additional qubits is
\begin{widetext}
\begin{align}
    \ket{\xi} &= 
        \alpha\ket*{W^{N-2}}\ket{\psi} + \beta\ket*{0^{\otimes N-2}}\ket{\phi} \nonumber\\
        &= 
        \alpha_{00}\ket*{W^{N-2}}\ket{00}
        + \alpha_{01}\ket*{W^{N-2}}\ket{01} 
        + \alpha_{10}\ket*{W^{N-2}}\ket{10}
        + \alpha_{11}\ket*{W^{N-2}}\ket{11} \nonumber\\
        &\quad 
        + \beta_{00}\ket*{0^{\otimes N-2}}\ket{00}
        + \beta_{01}\ket*{0^{\otimes N-2}}\ket{01} 
        + \beta_{10}\ket*{0^{\otimes N-2}}\ket{10}
        + \beta_{11}\ket*{0^{\otimes N-2}}\ket{11} 
    .
\end{align}
We can rewrite $\ket{\xi}$ by expanding $\ket*{W^k}$ as $\frac{1}{\sqrt{k}}\ket*{0^{\otimes k-1}}\ket{1} + \frac{\sqrt{k-1}}{\sqrt{k}}\ket*{W^{k-1}}\ket{0}$, yielding
\begin{align}
    \ket{\xi} &=
        \alpha_{00}\frac{1}{\sqrt{N-2}}\ket*{0^{\otimes N-3}}\ket{100} + \alpha_{00}\sqrt{\frac{N-3}{N-2}}\ket*{W^{N-3}}\ket{000} 
        + \beta_{00}\ket*{0^{\otimes N-3}}\ket{000} \nonumber\\
        &\quad+ \alpha_{01}\frac{1}{\sqrt{N-2}}\ket*{0^{\otimes N-3}}\ket{101} + \alpha_{01}\sqrt{\frac{N-3}{N-2}}\ket{W^{N-3}}\ket{001} 
        + \beta_{01}\ket*{0^{\otimes N-3}}\ket{001} \nonumber\\
        &\quad+ \alpha_{10}\frac{1}{\sqrt{N-2}}\ket*{0^{\otimes N-3}}\ket{110} + \alpha_{10}\sqrt{\frac{N-3}{N-2}}\ket{W^{N-3}}\ket{010}
        + \beta_{10}\ket*{0^{\otimes N-3}}\ket{010}\nonumber\\
        &\quad+ \alpha_{11}\frac{1}{\sqrt{N-2}}\ket*{0^{\otimes N-3}}\ket{111} + \alpha_{11}\sqrt{\frac{N-3}{N-2}}\ket{W^{N-3}}\ket{011}
        + \beta_{11}\ket*{0^{\otimes N-3}}\ket{011}
    .
\end{align}
Now, suppose we introduce one additional dissipator described by the jump operator $\LL_{\cbqty{N-2,N-1,N}}$. Since $\LL_{\cbqty{N-2,N-1,N}}$ only supports zero- or one-excitation stationary states, clearly $\ket{\xi}$ cannot be a stationary state unless we require $\alpha_{01} = \alpha_{10} = \alpha_{11} = \beta_{11} = 0$. This simplifies $\ket{\xi}$ to
\begin{align}
    \ket{\xi} &=
        \alpha_{00}\frac{\sqrt{N-3}}{\sqrt{N-2}}\ket*{W^{N-3}}\ket{000} \beta_{00}\ket*{0^{\otimes N-3}}\ket{000} \nonumber\\
        &\quad+ 
        \alpha_{00}\frac{1}{\sqrt{N-2}}\ket*{0^{\otimes N-3}}\ket{100} + \beta_{01}\ket*{0^{\otimes N-3}}\ket{001} + \beta_{10}\ket*{0^{\otimes N-3}}\ket{010} 
    .
\end{align}
Since $\LL_{jkl}\ket{000} = 0$, any value of $\beta_{00}$ can still be lead to a stationary state. On the other hand, $\beta_{01}$ and $\beta_{10}$ are constrained since only $\LL_{jkl}\ket*{W^3} = 0$. Thus we must require $\beta_{01} = \beta_{10} = \frac{\alpha_{00}}{\sqrt{N-2}}$, leading to 
\begin{align}
    \ket{\xi} &=
        \alpha_{00}\frac{1}{\sqrt{N-2}}\Big(\ket*{0^{\otimes N-3}}\ket{100} + \ket*{0^{\otimes N-3}}\ket{010} + \ket*{\otimes 0^{N-3}}\ket{001}\Big)
        + \alpha_{00}\sqrt{\frac{N-3}{N-2}}\ket*{W^{N-3}}\ket{000} \nonumber\\
        &\quad+ \beta_{00}\ket*{0^{N-3}}\ket{000} \nonumber\\
        &=
        \alpha_{00}\sqrt{\frac{N}{N-2}}\ket*{W^N} + \beta_{00}\ket*{0^{\otimes N}} \nonumber\\
        &=
        \alpha'\ket*{W^N} + \beta'\ket*{0^{\otimes N}}
        .
\end{align}
\end{widetext}
This is the most general pure stationary state of the new set of dissipators. Any odd $N$-qubit W state requires at minimum $\lfloor N/2\rfloor$ dissipators, and straightforward generalization shows that this is true for even $N$ as well. It can also be shown that selecting $\ket*{W^N}$ with width-$k$ dissipators requires at least $\lfloor N/(k-1) \rfloor$ dissipators.
%
%
\section{Ruling out other dark states}
\label{app:OtherStates}
%
The dissipator constructions we present in Sec.~\ref{sec:DissipatorConstruction} support several stationary states. When designing the Hamiltonian of our protocol, we ignore this fact and instead only consider the need for the Hamiltonian to distinguish between $\ket*{W^N}$ and the $\ket*{0^{\otimes N}}$. In this appendix, we will explicitly consider the other stationary states allowed by a particular dissipator construction and show that they do not cause a problem. The logic we use will easily extend to cover any $N$-qubit case, with any connected dissipator configuration.
\par
Let us consider the minimal connected width-3 5-qubit dissipator configuration as shown in Fig.~\ref{fig:diagrams}a, which is described by the two edges $E_1 = \cbqty{1, 2, 3}$ and $E_2 = \cbqty{3, 4, 5}$. If we choose the coefficients $\cbqty{1/\sqrt{6}, 1/\sqrt{6}, -2/\sqrt{6}}$ for $E_1$ and $\cbqty{-2/\sqrt{6}, 1/\sqrt{6}, 1/\sqrt{6}}$ for $E_2$, then it is straightforward to show that the kernel of the joint dissipation is spanned by the following five states:
\begin{subequations}
\begin{align}
    &\ket{00000} \\
    &\ket{W^5} \\
    &\ket{\Phi_1} \equiv \frac{1}{\sqrt{2}}\pqty{\ket{01000} - \ket{10000}} \\
    &\ket{\Phi_2} \equiv \frac{1}{\sqrt{2}}\pqty{\ket{00001} - \ket{00010}} \\
    &\ket{\Phi_3} \equiv \frac{1}{2}\pqty{\ket{01001} + \ket{10010} - \ket{01010} - \ket{10001}}
    .
\end{align}
\end{subequations}
By construction, we choose the parameters in the Hamiltonian $\HH$ from Eq.~\eqref{eqn:FinalH} such that $\ket{W^5}$ is an eigenstate. This will then be the only pure steady state of the combined Hamiltonian and dissipative dynamics provided there is no other eigenstate $\ket{\Psi}$ of $\HH$ in the kernel of the joint dissipation. That is, $\HH$ must \emph{not} support any state of the form
\begin{align}
    \ket{\Psi} \equiv 
        \alpha\ket{\Phi_1} + \beta\ket{\Phi_2} + \gamma\ket{\Phi_3} + \delta\ket{00000}
    ,
    \label{eqn:BadPsi}
\end{align}
as an eigenstate. 
\par
Suppose there were a $\ket{\Psi}$ such that $\HH\ket{\Psi} = \lambda\ket{\Psi}$. This eigenvalue equation leads to a set of 26 coupled equations in 5 unknowns (the four coefficients defining $\ket{\Psi}$ and the associated eigenvalue), as $\HH\ket{\Psi}$ can have support on any zero-, one-, two-, or three-excitation basis state. Such an over-determined set of constraint equations only has a solution if the parameters have been carefully tuned to support it, and so we expect such Hamiltonians to be exceedingly rare. For our purposes, it is straightforward to show that no state of the form in Eq.~\eqref{eqn:BadPsi} can be an eigenstate of any of the specific $5$-qubit Hamiltonians discussed in the main text. 
\par
Similar reasoning applies when considering other dissipator configurations. Generally, the complexity of our Hamiltonian means that the parameters would need to be carefully balanced to allow any stationary state in the kernel of the joint dissipation to be an eigenstate. The criteria listed at the end of Sec.~\ref{sec:HamiltonianConstruction} ensure that our Hamiltonians support $\ket*{W^N}$ as an eigenstate, it is extremely unlikely that we choose a set of parameters that \emph{also} satisfy the over-determined set of constraints necessary to allow another dark state to be an eigenstate. In fact, as $N$ grows we expect this to become even less likely, since the number of constraints coming from the eigenvalue equation grows extremely quickly. Our randomized simulations whose results were shown in Sec.~\ref{sec:Results} corroborate our expectation that failure is unlikely, as none of our simulations showed a failure to stabilize $\ket*{W^N}$, despite not taking this potential issue into account in their construction. 
\section{Insufficiency of bilinear interactions}
\label{app:Bilinear}
%
In this appendix, we show that it is not possible for a bilinear Hamiltonian of the form shown in Eq.~\eqref{eqn:BilinearH} to support $\ket*{W^N}$ as an eigenstate while \emph{not} supporting the state $\ket*{0^{\otimes N}}$ as an eigenstate. As discussed in the main text, since each term in $\HH^{(2)}$ always adds or removes two excitations, the $\ket*{W^N}$ could only be an eigenstate with eigenvalue zero, and only if $\mel*{jkl}{\HH^{(2)}}{W^N} = 0$ for every three-excitation state $\ket{jkl} \equiv \sigdag_j\sigdag_k\sigdag_l\ket*{0^{\otimes N}}$. Each of these states leads to a constraint on the coefficients entering $\HH^{(2)}$ of the form $c_{jk} + c_{jl} + c_{kl} = 0$, the collection of which can be represented simultaneously as a matrix equation
\begin{align}
    \mathbb{M}_N \vec{c} = 0
    ,
\end{align}
with $\vec{c} = \pmqty{c_{12} & c_{13} & \dots & c_{1N} & c_{23} & \dots & c_{NN}}^\intercal$ being the vector of coefficients and $\mathbb{M}_N$ containing the constraint information.
\par
There will be a non-trivial solution (i.e., a solution other than $\vec{c} = 0$) if the matrix $\mathbb{M}_N$ is not full rank. It has $\binom{N}{3}$ rows, one per three-excitation state, and $\binom{N}{2}$ columns, one per coefficient in $\HH$. We should generically expect that it is full rank (and so there is no non-trivial solution) whenever $\binom{N}{3} \ge \binom{N}{2}$, in other words when $N\ge5$. This counting argument is heuristic, and so the remainder of this appendix is dedicated to showing that $\mathbb{M}_N$ is full rank for all $N\ge5$.
\par
To begin, we can compute $\mathbb{M}_{4}$ and $\mathbb{M}_{5}$ explicitly and examine their structure
\begin{subequations}
\begin{align}
    \mathbb{M}_4
        &= \left[\begin{array}{c|cc|ccc}
            1&1&0&1&0&0\\
            1&0&1&0&1&0\\\cline{1-3}
            0&1&1&0&0&1\\\hline
            0&0&0&1&1&1
        \end{array}\right] \\
    \mathbb{M}_5
        &= \left[\begin{array}{c|ccc|cccccc}
        1&1&0&0&1&0&0&0&0&0\\ 
        1&0&1&0&0&1&0&0&0&0\\
        1&0&0&1&0&0&1&0&0&0\\ \cline{1-4}
        0&1&1&0&0&0&0&1&0&0\\ 
        0&1&0&1&0&0&0&0&1&0\\ 
        0&0&1&1&0&0&0&0&0&1\\ \hline
        0&0&0&0&1&1&0&1&0&0\\ 
        0&0&0&0&1&0&1&0&1&0\\ 
        0&0&0&0&0&1&1&0&0&1\\ 
        0&0&0&0&0&0&0&1&1&1
        \end{array}\right]
\end{align}
\end{subequations}
We have split these matrices into various blocks which illustrate the recursive structure of these matrices. Examining $\mathbb{M}_5$, there are two important features to note. 
Firstly, we focus on the top three rows. These rows correspond to the three-excitation states and associated constraints
\begin{subequations}
\begin{align}
    \mel{11100}{\HH^{(2)}}{W^N} &= c_{12} + c_{13} + c_{23} = 0 \\
    \mel{11010}{\HH^{(2)}}{W^N} &= c_{12} + c_{14} + c_{24} = 0 \\
    \mel{11001}{\HH^{(2)}}{W^N} &= c_{12} + c_{15} + c_{25} = 0 
    .
\end{align}
\end{subequations}
In general, there are $N-2$ states of the form $\sigdag_{j}\ket{110\dots0}$ which will lead to the first column of $\mathbb{M}_N$ consisting of $N-2$ ones followed by all zeros and to there being a block $\mathbb{I}_{N-2}$ immediately adjacent.
The second salient feature of $\mathbb{M}_5$ we can see is that $\mathbb{M}_4$ is included as the bottom-right block, padded with zeros. This block corresponds to constraints coming from states of the form $\ket{0}\ket{jkl}_4$, where all three excitations are in the rightmost qubits.
\par
Generalizing to any $N$, we can illustrate this structure schematically and construct $\mathbb{M}_{N+1}$ in terms of $\mathbb{M}_{N}$ as shown in Fig.~\ref{fig:constraintmatrix}.
\begin{figure}[h]
    \centering
    \begin{tikzpicture}[scale=0.9, every node/.style={transform shape}]
        \node at (0, 0) {\scalebox{0.7}{$
        \bigentry{\mathbb{M}_{N+1} =} 
        \left(\begin{BMAT}{c2c}{c2c}
        \begin{BMAT}{c2c}{c2c}
            \begin{BMAT}{c}{ccc}1\\1\\\rvdots\end{BMAT} & \begin{BMAT}(@,40pt,40pt){c}{c}\bigentry[\Large]{\mathbb{I}_{N-1}}\end{BMAT} \\
            \bigentry[\Large]{{0}} & \begin{BMAT}(@,40pt,60pt){c}{c}\bigentry[\Large]{\mathbb{P}_{N}}\end{BMAT} 
        \end{BMAT} & \begin{BMAT}(@,100pt,100pt){c}{c}\bigentry{\mathbb{I}_{\binom{N}{3}}}\end{BMAT}\\ 
        \bigentry{{0}} & \begin{BMAT}(@,100pt,140pt){c}{c}\bigentry{\mathbb{M}_{N}}\end{BMAT}
        \end{BMAT}\right)
        $}};
        
        \draw [decorate, decoration = {brace,mirror}, thick] (3.2, 2.05) --  (3.2,3.15);
        \draw [decorate, decoration = {brace,mirror}, thick] (3.2, 0.42) --  (3.2,1.99);
        \draw [decorate, decoration = {brace,mirror}, thick] (3.2,-3.13) --  (3.2,0.36);
        
        \node at (4.0,  2.600) {$\ket{11\cdots}$};
        \node at (4.0,  1.205) {$\ket{10\cdots}$};
        \node at (4.0, -1.385) {$\ket{0\cdot\cdots}$};
    \end{tikzpicture}
    \caption{Illustration of the structure of $\mathbb{M}_{N+1}$ in terms of the previous $\mathbb{M}_{N}$. $\mathbb{P}_{N+1}$ is the full top-left block, though its structure is generally unimportant.}
    \label{fig:constraintmatrix}
\end{figure}
\par
Now, we will show that if $\mathbb{M}_N$ is full-rank, then all the columns of $\mathbb{M}_{N+1}$ are linearly independent and so it is full-rank as well. To start, we can ignore the rightmost $\binom{N}{2}$ columns. By our assumption all the columns of $\mathbb{M}_{N}$ are linearly independent, and since all other entries in those rows of $\mathbb{M}_{N+1}$ are zero \emph{all} these columns are linearly independent of the remaining columns of $\mathbb{M}_{N+1}$.
What remains is to show that the first $N$ columns of $\mathbb{M}_{N+1}$ are linearly independent. Clearly the columns corresponding to the $\mathbb{I}_{N-1}$ block form a linearly independent set. Since the first $N-1$ entries of the first column are all $1$, if we were to write the first column in terms of the next $N-1$ columns, then each coefficient would have to be $1$. However, it is straightforward to see that the full first column is not equal to the sum of the next $N-1$.
Therefore, all the columns of $\mathbb{M}_{N+1}$ form a linearly independent set and so $\mathbb{M}_{N+1}$ is full-rank, provided $\mathbb{M}_{N}$ is.
\par
Finally, an explicit calculation shows that $\mathbb{M}_{5}$ is full-rank, leading us to conclude that $\mathbb{M}_N$ has full rank whenever $k\ge5$ and therefore in that case there is no nontrivial choice of coefficients which can allow the bilinear Hamiltonian shown in Eq.~\eqref{eqn:BilinearH} to support $\ket*{W^N}$ as an eigenstate. 
\par
This same procedure can be applied to to show that, for a Hamiltonian consisting of $p$-qubit excitation operators of the form $\sigdag_{j_1}\sigdag_{j_2}\dots\sigdag_{j_p}$ (and their hermitian conjugate), there is no solution whenever $N\ge2p+1$. Instead, as discussed in Sec.~ \ref{sec:HamiltonianConstruction}, we must consider generalized hopping interactions on three or more qubits, with each term including both raising and lowering operators.
%
\section{Stabilization with width-4 configurations}
\label{app:Width4}
%
\begin{figure}[t!]
    \centering
    \includegraphics[width=\columnwidth]{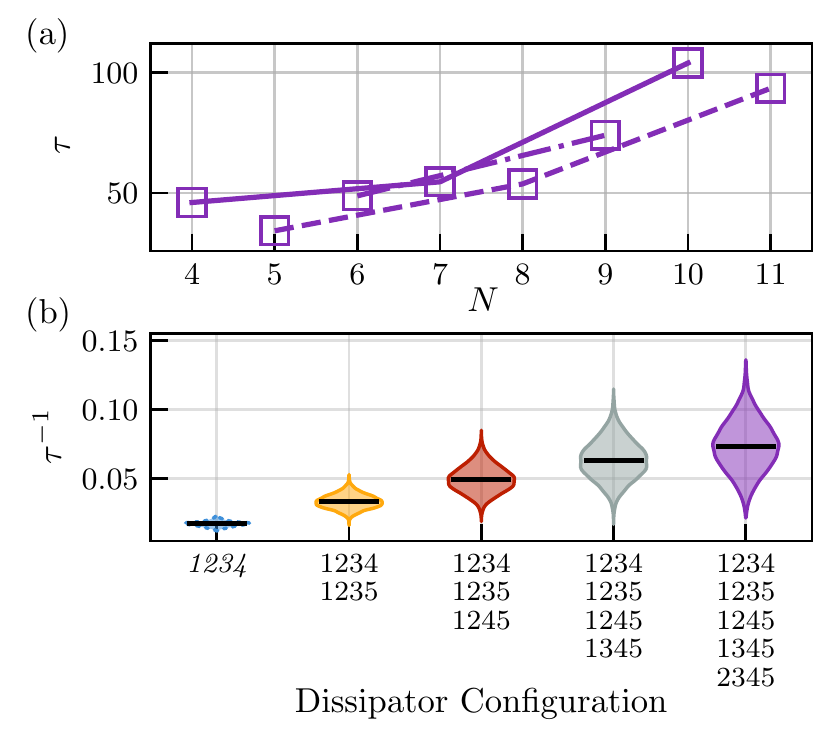}
    \caption{
    (a) Scaling of the stabilization time constant with the number of qubits $N$, when using the maximal Hamiltonian and the minimal width-$4$ dissipator configurations discussed in the text. The time constants are only slightly longer than those shown in purple in Fig.~\ref{fig:scaling} of the main text, which used the same maximal Hamiltonian and minimal width-$3$ configurations. 
    (b) Distributions of stabilization rates for the 5 distinct configurations of width-$4$ dissipators for $N=5$. Each simulation was run 10240 times, with the coefficients of each dissipator chosen randomly for each run according to a procedure analogous to that presented in Sec.~\ref{ssec:CompareConfigs}. The median of each distribution is marked with a black bar.
    }
    \label{fig:Width4}
\end{figure}
When using width-$3$ dissipators, we found that it was necessary to use $\lfloor N/2\rfloor$ dissipators to have a connected configuration. We then chose to use the chain/ring configurations to examine the scaling of our protocol, as these configurations are easily extended to $N$ qubits where $N$ is odd/even, respectively. For width-$4$ dissipators, minimal configurations have $\lceil(N-1)/3\rceil$ dissipators. We use the set of hyperedges,
\begin{align}
    E_1 &= \{1, 2, 3, 4\} \nonumber\\ 
    E_2 &= \{4, 5, 6, 7\} \nonumber\\
        &\cdots \nonumber\\ 
    E_{\lceil(N-1)/3\rceil} &= \{N-3, N-2, N-1, N\}
    ,
\end{align}
to define the dissipator configuration for $N$ qubits. When $N\equiv1\mod3$, this is a straightforward generalization of the width-$3$ chain configuration. The other two cases are only slightly different. Figure~\ref{fig:Width4}a shows the scaling of the stabilization time constant with $N$ for the maximal Hamiltonian paired with these width-$4$ configurations, with the dissipator coefficients chosen arbitrarily as
\begin{alignat}{2}
    r_1[E] &= 0.584
    &\quad
    r_2[E] &= -0.146 + 0.545i
    \nonumber \\
    r_3[E] &= -0.438 - 0.253i
    &\quad
    r_4[E] &= -0.292i
    .
\end{alignat}
Note that while stabilization is slower than observed with width-$3$ dissipator configurations, $\tau$ still appears to be scaling polynomially with $N$ rather than exponentially as observed with a single global dissipator.
\par
As a comparison to the results presented in Sec.~\ref{ssec:CompareConfigs}, Fig.~\ref{fig:Width4}b shows the distribution of stabilization rates obtained when simulating the five distinct configurations of width-$4$ dissipators on $N=5$ qubits, with randomized parameters. Note that unlike the width-$3$ case, here the median performance does scale monotonically with the number of dissipators. Presumably, repeating these simulations for all width-$3$ and width-$4$ configurations for some $N\gg3,4$ would yield similar behavior. Unfortunately, both the number of configurations and the simulation time required grow too quickly for such a comparison to be tractable in this work.
%
\section{Enumerating all 5-qubit width-3 configurations}
\label{app:EnumerateFive}
%
In this appendix, we enumerate all the distinct configurations of width-3 dissipators on 5 qubits with the maximal Hamiltonian, and discuss how this procedure generalizes to different scenarios. To begin, we must first examine what it means for two dissipator configurations to be ``distinct''.
\par
Consider a version of our protocol employing the maximal Hamiltonian with the coefficients derived from the $N$'th roots of unity as listed in Eq.~\eqref{eqn:CoeffsMax}, along with some configuration of dissipators described by $\{E_j\}$ with $E_j = \{q^1_j, q^2_j, \dots\}$ and some set of coefficients $r_k[E_j]$ for each jump operator. We find that if we choose some permutation of the qubit indices $p$, then the performance characteristics of a second configuration described by a different set of jump operators defined by $\{\tilde{E}_j\}$ where $\tilde{E}_j = \{p(q^1_j), p(q^2_j), \dots\}$ and the same coefficients $r_k[\tilde{E}_j] = r_k[E_j]$ are the same as the original configuration. Note that we are only permuting the qubit indices in the jump operators, \emph{not} the Hamiltonian. We therefore are led to identify two dissipator configurations which can be related by a permutation of the qubit indices as being two representatives of a single class of equivalent configurations. This is possible since, in some sense, our symmetric parameter choice leads the Hamiltonian itself to treat each qubit symmetrically. In contrast, the minimal Hamiltonian is less symmetric, and a protocol employing it would lead us to identify dissipator configurations which are related by a restricted class of permutations which do not mix the first two qubits with the remaining $N-2$. In this appendix we consider only the maximal Hamiltonian with symmetric coefficients as this leads to the smallest set of distinct dissipator configurations.
\par
The hypergraph representing a configuration built entirely from width-$3$ dissipators is a $3$-uniform hypergraph. Enumerating all distinct configurations on $N$-qubits is then equivalent to enumerating all $3$-uniform hypergraphs on $N$ unlabelled vertices. Such a hypergraph can be turned into a dissipator configuration by arbitrarily labelling each vertex with a qubit index. Note that we exclude the empty graph (which would correspond to no dissipation at all), so there will be one less configuration than possible hypergraphs. For $N=5$ qubits, there are 33 distinct dissipator configurations as listed in Table~\ref{tab:Configs}.
\begin{table}[t]
    \centering
    \begin{tabular}{|l|l|}
        \hline
        $m=1$\Tstrut\Bstrut  & $\mathit{\{123\}}$ \\
        \hline
        $m=2$\Tstrut\Bstrut  & $\mathit{\{123, 124\}}$, $\mathbf{\{123, 145\}}$ \\ 
        \hline
        \multirow{2}{3.5em}{$m=3$}  
        & $\mathit{\{123, 124, 134\}}$, $\{123, 124, 125\}$, $\{123, 124, 135\}$,\Tstrut \\
        & $\{123, 124, 345\}$\Bstrut \\
        \hline
        \multirow{3}{3.5em}{$m=4$}  
        & $\{123, 124, 125, 134\}$, $\{123, 124, 125, 345\}$,\Tstrut \\ 
        & $\{123, 124, 134, 234\}$, $\{123, 124, 134, 235\}$, \\ 
        & $\{123, 124, 135, 145\}$, $\{123, 124, 135, 245\}$\Bstrut \\
        \hline
        \multirow{3}{3.5em}{$m=5$}  
        & $\{123, 124, 125, 134, 135\}$, $\{123, 124, 125, 134, 234\}$,\Tstrut \\
        & $\{123, 124, 125, 134, 235\}$, $\{123, 124, 125, 134, 345\}$, \\
        & $\{123, 124, 134, 235, 245\}$, $\{123, 124, 135, 245, 345\}$\Bstrut \\
        \hline
        \multirow{6}{3.5em}{$m=6$} 
        & $\{123, 124, 125, 134, 135, 145\}$,\Tstrut \\
        & $\{123, 124, 125, 134, 135, 234\}$, \\
        & $\{123, 124, 125, 134, 135, 245\}$, \\
        & $\{123, 124, 125, 134, 234, 345\}$, \\
        & $\{123, 124, 125, 134, 235, 345\}$, \\
        & $\{123, 124, 134, 235, 245, 345\}$\Bstrut \\
        \hline
        \multirow{4}{3.5em}{$m=7$}  
        & $\{123, 124, 125, 134, 135, 145, 234\}$,\Tstrut \\
        & $\{123, 124, 125, 134, 135, 234, 235\}$, \\
        & $\{123, 124, 125, 134, 135, 234, 245\}$, \\
        & $\{123, 124, 125, 134, 135, 245, 345\}$\Bstrut \\
        \hline
        \multirow{2}{3.5em}{$m=8$}
        & $\{123, 124, 125, 134, 135, 145, 234, 235\}$,\Tstrut \\
        & $\{123, 124, 125, 134, 135, 234, 245, 345\}$\Bstrut \\
        \hline
        $m=9$\Tstrut\Bstrut  & $\{123, 124, 125, 134, 135, 145, 234, 235, 245\}$ \\
        \hline
        $m=10$\Tstrut\Bstrut & $\{123, 124, 125, 134, 135, 145, 234, 235, 245, 345\}$ \\
        \hline
    \end{tabular}
    \caption{List of distinct (i.e. unique up to relabelling of the qubits) configurations of width-3 dissipators on 5 qubits, denoted as a set of hyperedges. To conserve space we write each hyperedge as a list of numbers rather than as a set, e.g. we write $123$ for $\{1, 2, 3\}$. Disconnected configurations are italicized, and the unique minimal connected configuration is shown in bold. The order in which the configurations are listed matches the order used on the horizontal axis of Fig.~\ref{fig:5qbConfigs}.}
    \label{tab:Configs}
\end{table}
\par
The procedure followed to enumerate the distinct width-$w$ configurations on $N$ qubits subject to the symmetric maximal Hamiltonian is the same, each configuration corresponds to some arbitrary labelling of a $w$-uniform hypergraph on $N$ unlabelled vertices. If instead the minimal Hamiltonian is employed, each distinct configuration corresponds to a $w$-uniform hypergraph on $N$ vertices, where two vertices are colored. Arbitrarily assigning the indices 0 and 1 to the two colored vertices and the remaining indices to the remaining vertices produces a representative dissipator configuration. The procedure is similar for other Hamiltonians: dissipator configurations correspond to arbitrary labellings of $w$-uniform hypergraphs on $N$ vertices, possibly subject to some asymmetry depending on the details of the Hamiltonian. Two configurations are distinct if they are labellings of different hypergraphs.
%
\section{Effect of asymmetric local decoherence rates}
\label{app:GammaDist}
%
\begin{figure}[t!]
    \centering
    \includegraphics[width=\columnwidth]{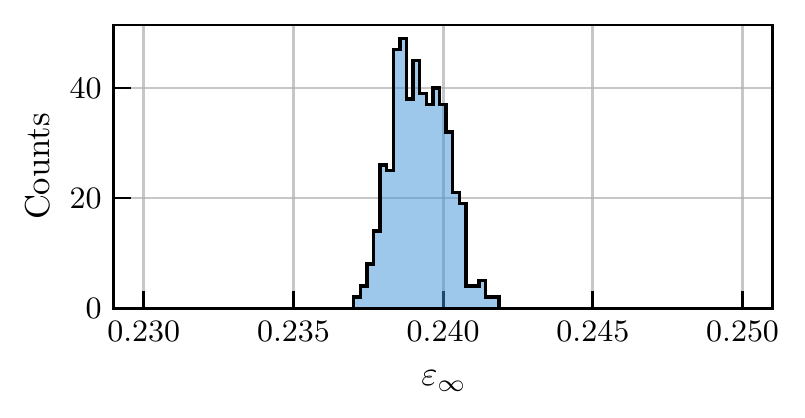}
    \caption{Histogram of steady-state errors from simulations of Eq.~\eqref{eqn:SchemeDecayLV} with $500$ randomly chosen sets of decoherence rates $\{\gamma^{-}_j,\gamma^{z}_j\}$, with $\gamma^W_\downarrow$ fixed. The sharp distribution indicates that the details of the local decoherence rates are not important.}
    \label{fig:DecayDist}
\end{figure}
Based on the simple rate model presented in Sec.~\ref{ssec:TargetState}, we expect that any set of decoherence rates which leads to the same effective decay rate $\gamma^W_\downarrow$ will yield nearly the same steady-state error $\varepsilon_\infty$, set by the competition between the stabilization rate and the decay rate. To test this, we simulated the stabilization of $\ket*{W^5}$ with the version of our protocol employing the maximal Hamiltonian and two width-$3$ dissipators in the configuration displayed in Fig.~\ref{fig:diagrams}, including local relaxation and dephasing processes on each qubit. We performed $500$ simulations, each with randomly selected decoherence rates $\{\gamma^{-}_j,\gamma^{z}_j\}$ for each qubit, constrained to yield the same $\gamma^W_\downarrow = \lambda\times10^{-2}$. As shown in Fig.~\ref{fig:DecayDist}, the distribution of $\varepsilon_\infty$ is tight, with all values contained within $0.239\pm0.003$. From this, we conclude that the distribution of bare decoherence rates have little impact on our protocol, since the effective decay rate of the target state is formed by averaging the decoherence rates of each qubit as shown in Eq.~\eqref{eqn:GenericDecayRate}.
%
%
%

\begin{thebibliography}{61}
\expandafter\ifx\csname natexlab\endcsname\relax\def\natexlab#1{#1}\fi
\expandafter\ifx\csname bibnamefont\endcsname\relax
  \def\bibnamefont#1{#1}\fi
\expandafter\ifx\csname bibfnamefont\endcsname\relax
  \def\bibfnamefont#1{#1}\fi
\expandafter\ifx\csname citenamefont\endcsname\relax
  \def\citenamefont#1{#1}\fi
\expandafter\ifx\csname url\endcsname\relax
  \def\url#1{\texttt{#1}}\fi
\expandafter\ifx\csname urlprefix\endcsname\relax\def\urlprefix{URL }\fi
\providecommand{\bibinfo}[2]{#2}
\providecommand{\eprint}[2][]{\url{#2}}

\bibitem[{\citenamefont{Raussendorf and Briegel}(2001)}]{Raussendorf2001}
\bibinfo{author}{\bibfnamefont{R.}~\bibnamefont{Raussendorf}} \bibnamefont{and}
  \bibinfo{author}{\bibfnamefont{H.~J.} \bibnamefont{Briegel}},
  \bibinfo{journal}{Phys. Rev. Lett.} \textbf{\bibinfo{volume}{86}},
  \bibinfo{pages}{5188} (\bibinfo{year}{2001}),
  \urlprefix\url{https://link.aps.org/doi/10.1103/PhysRevLett.86.5188}.

\bibitem[{\citenamefont{Jozsa and Linden}(2003)}]{Jozsa2003}
\bibinfo{author}{\bibfnamefont{R.}~\bibnamefont{Jozsa}} \bibnamefont{and}
  \bibinfo{author}{\bibfnamefont{N.}~\bibnamefont{Linden}},
  \bibinfo{journal}{Proceedings: Mathematical, Physical and Engineering
  Sciences} \textbf{\bibinfo{volume}{459}}, \bibinfo{pages}{2011}
  (\bibinfo{year}{2003}), \urlprefix\url{http://www.jstor.org/stable/3560059}.

\bibitem[{\citenamefont{Degen et~al.}(2017)\citenamefont{Degen, Reinhard, and
  Cappellaro}}]{Cappellaro2017}
\bibinfo{author}{\bibfnamefont{C.~L.} \bibnamefont{Degen}},
  \bibinfo{author}{\bibfnamefont{F.}~\bibnamefont{Reinhard}}, \bibnamefont{and}
  \bibinfo{author}{\bibfnamefont{P.}~\bibnamefont{Cappellaro}},
  \bibinfo{journal}{Rev. Mod. Phys.} \textbf{\bibinfo{volume}{89}},
  \bibinfo{pages}{035002} (\bibinfo{year}{2017}),
  \urlprefix\url{https://link.aps.org/doi/10.1103/RevModPhys.89.035002}.

\bibitem[{\citenamefont{Cirac et~al.}(1999)\citenamefont{Cirac, Ekert, Huelga,
  and Macchiavello}}]{Cirac1999}
\bibinfo{author}{\bibfnamefont{J.~I.} \bibnamefont{Cirac}},
  \bibinfo{author}{\bibfnamefont{A.~K.} \bibnamefont{Ekert}},
  \bibinfo{author}{\bibfnamefont{S.~F.} \bibnamefont{Huelga}},
  \bibnamefont{and}
  \bibinfo{author}{\bibfnamefont{C.}~\bibnamefont{Macchiavello}},
  \bibinfo{journal}{Phys. Rev. A} \textbf{\bibinfo{volume}{59}},
  \bibinfo{pages}{4249} (\bibinfo{year}{1999}),
  \urlprefix\url{https://link.aps.org/doi/10.1103/PhysRevA.59.4249}.

\bibitem[{\citenamefont{Gisin and Thew}(2007)}]{Gisin2007}
\bibinfo{author}{\bibfnamefont{N.}~\bibnamefont{Gisin}} \bibnamefont{and}
  \bibinfo{author}{\bibfnamefont{R.}~\bibnamefont{Thew}},
  \bibinfo{journal}{Nature Photonics} \textbf{\bibinfo{volume}{1}},
  \bibinfo{pages}{165} (\bibinfo{year}{2007}).

\bibitem[{\citenamefont{Pirandola et~al.}(2020)\citenamefont{Pirandola,
  Andersen, Banchi, Berta, Bunandar, Colbeck, Englund, Gehring, Lupo, Ottaviani
  et~al.}}]{Pirandola2020}
\bibinfo{author}{\bibfnamefont{S.}~\bibnamefont{Pirandola}},
  \bibinfo{author}{\bibfnamefont{U.~L.} \bibnamefont{Andersen}},
  \bibinfo{author}{\bibfnamefont{L.}~\bibnamefont{Banchi}},
  \bibinfo{author}{\bibfnamefont{M.}~\bibnamefont{Berta}},
  \bibinfo{author}{\bibfnamefont{D.}~\bibnamefont{Bunandar}},
  \bibinfo{author}{\bibfnamefont{R.}~\bibnamefont{Colbeck}},
  \bibinfo{author}{\bibfnamefont{D.}~\bibnamefont{Englund}},
  \bibinfo{author}{\bibfnamefont{T.}~\bibnamefont{Gehring}},
  \bibinfo{author}{\bibfnamefont{C.}~\bibnamefont{Lupo}},
  \bibinfo{author}{\bibfnamefont{C.}~\bibnamefont{Ottaviani}},
  \bibnamefont{et~al.}, \bibinfo{journal}{Adv. Opt. Photon.}
  \textbf{\bibinfo{volume}{12}}, \bibinfo{pages}{1012} (\bibinfo{year}{2020}),
  \urlprefix\url{https://opg.optica.org/aop/abstract.cfm?URI=aop-12-4-1012}.

\bibitem[{\citenamefont{Preskill}(2012)}]{Preskill2012}
\bibinfo{author}{\bibfnamefont{J.}~\bibnamefont{Preskill}},
  \emph{\bibinfo{title}{Quantum computing and the entanglement frontier}}
  (\bibinfo{year}{2012}), \urlprefix\url{https://arxiv.org/abs/1203.5813}.

\bibitem[{\citenamefont{Terhal}(2015)}]{Terhal2015}
\bibinfo{author}{\bibfnamefont{B.~M.} \bibnamefont{Terhal}},
  \bibinfo{journal}{Rev. Mod. Phys.} \textbf{\bibinfo{volume}{87}},
  \bibinfo{pages}{307} (\bibinfo{year}{2015}),
  \urlprefix\url{https://link.aps.org/doi/10.1103/RevModPhys.87.307}.

\bibitem[{\citenamefont{Georgescu}(2020)}]{Georgescu2020}
\bibinfo{author}{\bibfnamefont{I.}~\bibnamefont{Georgescu}},
  \bibinfo{journal}{Nature Reviews Physics} \textbf{\bibinfo{volume}{2}},
  \bibinfo{pages}{519} (\bibinfo{year}{2020}),
  \urlprefix\url{https://doi.org/10.1038/s42254-020-0244-y}.

\bibitem[{\citenamefont{Lidar}(2014)}]{Lidar2014}
\bibinfo{author}{\bibfnamefont{D.~A.} \bibnamefont{Lidar}},
  \emph{\bibinfo{title}{Review of Decoherence-Free Subspaces, Noiseless
  Subsystems, and Dynamical Decoupling}} (\bibinfo{publisher}{John Wiley \&
  Sons, Ltd}, \bibinfo{year}{2014}), pp. \bibinfo{pages}{295--354}, ISBN
  \bibinfo{isbn}{9781118742631},
  \urlprefix\url{https://onlinelibrary.wiley.com/doi/abs/10.1002/9781118742631.ch11}.

\bibitem[{\citenamefont{Suter and \'Alvarez}(2016)}]{Suter2016}
\bibinfo{author}{\bibfnamefont{D.}~\bibnamefont{Suter}} \bibnamefont{and}
  \bibinfo{author}{\bibfnamefont{G.~A.} \bibnamefont{\'Alvarez}},
  \bibinfo{journal}{Rev. Mod. Phys.} \textbf{\bibinfo{volume}{88}},
  \bibinfo{pages}{041001} (\bibinfo{year}{2016}),
  \urlprefix\url{https://link.aps.org/doi/10.1103/RevModPhys.88.041001}.

\bibitem[{\citenamefont{Temme et~al.}(2017)\citenamefont{Temme, Bravyi, and
  Gambetta}}]{Temme2017}
\bibinfo{author}{\bibfnamefont{K.}~\bibnamefont{Temme}},
  \bibinfo{author}{\bibfnamefont{S.}~\bibnamefont{Bravyi}}, \bibnamefont{and}
  \bibinfo{author}{\bibfnamefont{J.~M.} \bibnamefont{Gambetta}},
  \bibinfo{journal}{Phys. Rev. Lett.} \textbf{\bibinfo{volume}{119}},
  \bibinfo{pages}{180509} (\bibinfo{year}{2017}),
  \urlprefix\url{https://link.aps.org/doi/10.1103/PhysRevLett.119.180509}.

\bibitem[{\citenamefont{Endo et~al.}(2018)\citenamefont{Endo, Benjamin, and
  Li}}]{Endo2018}
\bibinfo{author}{\bibfnamefont{S.}~\bibnamefont{Endo}},
  \bibinfo{author}{\bibfnamefont{S.~C.} \bibnamefont{Benjamin}},
  \bibnamefont{and} \bibinfo{author}{\bibfnamefont{Y.}~\bibnamefont{Li}},
  \bibinfo{journal}{Phys. Rev. X} \textbf{\bibinfo{volume}{8}},
  \bibinfo{pages}{031027} (\bibinfo{year}{2018}),
  \urlprefix\url{https://link.aps.org/doi/10.1103/PhysRevX.8.031027}.

\bibitem[{\citenamefont{O'Brien et~al.}(2021)\citenamefont{O'Brien, Polla,
  Rubin, Huggins, McArdle, Boixo, McClean, and Babbush}}]{OBrien2021}
\bibinfo{author}{\bibfnamefont{T.~E.} \bibnamefont{O'Brien}},
  \bibinfo{author}{\bibfnamefont{S.}~\bibnamefont{Polla}},
  \bibinfo{author}{\bibfnamefont{N.~C.} \bibnamefont{Rubin}},
  \bibinfo{author}{\bibfnamefont{W.~J.} \bibnamefont{Huggins}},
  \bibinfo{author}{\bibfnamefont{S.}~\bibnamefont{McArdle}},
  \bibinfo{author}{\bibfnamefont{S.}~\bibnamefont{Boixo}},
  \bibinfo{author}{\bibfnamefont{J.~R.} \bibnamefont{McClean}},
  \bibnamefont{and} \bibinfo{author}{\bibfnamefont{R.}~\bibnamefont{Babbush}},
  \bibinfo{journal}{PRX Quantum} \textbf{\bibinfo{volume}{2}},
  \bibinfo{pages}{020317} (\bibinfo{year}{2021}),
  \urlprefix\url{https://link.aps.org/doi/10.1103/PRXQuantum.2.020317}.

\bibitem[{\citenamefont{Koczor}(2021)}]{Koczor2021}
\bibinfo{author}{\bibfnamefont{B.}~\bibnamefont{Koczor}},
  \bibinfo{journal}{Phys. Rev. X} \textbf{\bibinfo{volume}{11}},
  \bibinfo{pages}{031057} (\bibinfo{year}{2021}),
  \urlprefix\url{https://link.aps.org/doi/10.1103/PhysRevX.11.031057}.

\bibitem[{\citenamefont{Verstraete et~al.}(2009)\citenamefont{Verstraete, Wolf,
  and Cirac}}]{Verstraete2009}
\bibinfo{author}{\bibfnamefont{F.}~\bibnamefont{Verstraete}},
  \bibinfo{author}{\bibfnamefont{M.~M.} \bibnamefont{Wolf}}, \bibnamefont{and}
  \bibinfo{author}{\bibfnamefont{J.~I.} \bibnamefont{Cirac}},
  \bibinfo{journal}{Nat. Phys.} \textbf{\bibinfo{volume}{5}},
  \bibinfo{pages}{633} (\bibinfo{year}{2009}),
  \urlprefix\url{http://dx.doi.org/10.1038/nphys1342}.

\bibitem[{\citenamefont{Harrington et~al.}(2022)\citenamefont{Harrington,
  Mueller, and Murch}}]{Harrington2022}
\bibinfo{author}{\bibfnamefont{P.~M.} \bibnamefont{Harrington}},
  \bibinfo{author}{\bibfnamefont{E.~J.} \bibnamefont{Mueller}},
  \bibnamefont{and} \bibinfo{author}{\bibfnamefont{K.~W.} \bibnamefont{Murch}},
  \bibinfo{journal}{Nature Reviews Physics} \textbf{\bibinfo{volume}{4}},
  \bibinfo{pages}{660} (\bibinfo{year}{2022}),
  \urlprefix\url{https://doi.org/10.1038/s42254-022-00494-8}.

\bibitem[{\citenamefont{Doucet et~al.}(2020)\citenamefont{Doucet, Reiter,
  Ranzani, and Kamal}}]{Doucet20}
\bibinfo{author}{\bibfnamefont{E.}~\bibnamefont{Doucet}},
  \bibinfo{author}{\bibfnamefont{F.}~\bibnamefont{Reiter}},
  \bibinfo{author}{\bibfnamefont{L.}~\bibnamefont{Ranzani}}, \bibnamefont{and}
  \bibinfo{author}{\bibfnamefont{A.}~\bibnamefont{Kamal}},
  \bibinfo{journal}{Phys. Rev. Research} \textbf{\bibinfo{volume}{2}},
  \bibinfo{pages}{023370} (\bibinfo{year}{2020}),
  \urlprefix\url{https://link.aps.org/doi/10.1103/PhysRevResearch.2.023370}.

\bibitem[{\citenamefont{Brown et~al.}(2021)\citenamefont{Brown, Doucet, Ristè,
  Ribeill, Cicak, Aumentado, Simmonds, Govia, Kamal, and Ranzani}}]{Brown21}
\bibinfo{author}{\bibfnamefont{T.}~\bibnamefont{Brown}},
  \bibinfo{author}{\bibfnamefont{E.}~\bibnamefont{Doucet}},
  \bibinfo{author}{\bibfnamefont{D.}~\bibnamefont{Ristè}},
  \bibinfo{author}{\bibfnamefont{G.}~\bibnamefont{Ribeill}},
  \bibinfo{author}{\bibfnamefont{K.}~\bibnamefont{Cicak}},
  \bibinfo{author}{\bibfnamefont{J.}~\bibnamefont{Aumentado}},
  \bibinfo{author}{\bibfnamefont{R.}~\bibnamefont{Simmonds}},
  \bibinfo{author}{\bibfnamefont{L.}~\bibnamefont{Govia}},
  \bibinfo{author}{\bibfnamefont{A.}~\bibnamefont{Kamal}}, \bibnamefont{and}
  \bibinfo{author}{\bibfnamefont{L.}~\bibnamefont{Ranzani}}
  (\bibinfo{year}{2021}), \eprint{arXiv:2107.13579},
  \urlprefix\url{https://arxiv.org/abs/2107.13579}.

\bibitem[{\citenamefont{Kimchi-Schwartz
  et~al.}(2016)\citenamefont{Kimchi-Schwartz, Martin, Flurin, Aron, Kulkarni,
  T\"ureci, and Siddiqi}}]{Siddiqi2016}
\bibinfo{author}{\bibfnamefont{M.~E.} \bibnamefont{Kimchi-Schwartz}},
  \bibinfo{author}{\bibfnamefont{L.}~\bibnamefont{Martin}},
  \bibinfo{author}{\bibfnamefont{E.}~\bibnamefont{Flurin}},
  \bibinfo{author}{\bibfnamefont{C.}~\bibnamefont{Aron}},
  \bibinfo{author}{\bibfnamefont{M.}~\bibnamefont{Kulkarni}},
  \bibinfo{author}{\bibfnamefont{H.~E.} \bibnamefont{T\"ureci}},
  \bibnamefont{and} \bibinfo{author}{\bibfnamefont{I.}~\bibnamefont{Siddiqi}},
  \bibinfo{journal}{Phys. Rev. Lett.} \textbf{\bibinfo{volume}{116}},
  \bibinfo{pages}{240503} (\bibinfo{year}{2016}),
  \urlprefix\url{https://link.aps.org/doi/10.1103/PhysRevLett.116.240503}.

\bibitem[{\citenamefont{Shankar et~al.}(2013)\citenamefont{Shankar, Hatridge,
  Leghtas, Sliwa, Narla, Vool, Girvin, Frunzio, Mirrahimi, and
  Devoret}}]{Shankar2013}
\bibinfo{author}{\bibfnamefont{S.}~\bibnamefont{Shankar}},
  \bibinfo{author}{\bibfnamefont{M.}~\bibnamefont{Hatridge}},
  \bibinfo{author}{\bibfnamefont{Z.}~\bibnamefont{Leghtas}},
  \bibinfo{author}{\bibfnamefont{K.}~\bibnamefont{Sliwa}},
  \bibinfo{author}{\bibfnamefont{A.}~\bibnamefont{Narla}},
  \bibinfo{author}{\bibfnamefont{U.}~\bibnamefont{Vool}},
  \bibinfo{author}{\bibfnamefont{S.~M.} \bibnamefont{Girvin}},
  \bibinfo{author}{\bibfnamefont{L.}~\bibnamefont{Frunzio}},
  \bibinfo{author}{\bibfnamefont{M.}~\bibnamefont{Mirrahimi}},
  \bibnamefont{and} \bibinfo{author}{\bibfnamefont{M.~H.}
  \bibnamefont{Devoret}}, \bibinfo{journal}{Nature}
  \textbf{\bibinfo{volume}{504}}, \bibinfo{pages}{419} (\bibinfo{year}{2013}).

\bibitem[{\citenamefont{Aron et~al.}(2014)\citenamefont{Aron, Kulkarni, and
  T\"ureci}}]{Tureci2014}
\bibinfo{author}{\bibfnamefont{C.}~\bibnamefont{Aron}},
  \bibinfo{author}{\bibfnamefont{M.}~\bibnamefont{Kulkarni}}, \bibnamefont{and}
  \bibinfo{author}{\bibfnamefont{H.~E.} \bibnamefont{T\"ureci}},
  \bibinfo{journal}{Phys. Rev. A} \textbf{\bibinfo{volume}{90}},
  \bibinfo{pages}{062305} (\bibinfo{year}{2014}),
  \urlprefix\url{https://link.aps.org/doi/10.1103/PhysRevA.90.062305}.

\bibitem[{\citenamefont{Cole et~al.}(2022)\citenamefont{Cole, Erickson,
  Zarantonello, Horn, Hou, Wu, Slichter, Reiter, Koch, and
  Leibfried}}]{Liebfried2022}
\bibinfo{author}{\bibfnamefont{D.~C.} \bibnamefont{Cole}},
  \bibinfo{author}{\bibfnamefont{S.~D.} \bibnamefont{Erickson}},
  \bibinfo{author}{\bibfnamefont{G.}~\bibnamefont{Zarantonello}},
  \bibinfo{author}{\bibfnamefont{K.~P.} \bibnamefont{Horn}},
  \bibinfo{author}{\bibfnamefont{P.-Y.} \bibnamefont{Hou}},
  \bibinfo{author}{\bibfnamefont{J.~J.} \bibnamefont{Wu}},
  \bibinfo{author}{\bibfnamefont{D.~H.} \bibnamefont{Slichter}},
  \bibinfo{author}{\bibfnamefont{F.}~\bibnamefont{Reiter}},
  \bibinfo{author}{\bibfnamefont{C.~P.} \bibnamefont{Koch}}, \bibnamefont{and}
  \bibinfo{author}{\bibfnamefont{D.}~\bibnamefont{Leibfried}},
  \bibinfo{journal}{Phys. Rev. Lett.} \textbf{\bibinfo{volume}{128}},
  \bibinfo{pages}{080502} (\bibinfo{year}{2022}),
  \urlprefix\url{https://link.aps.org/doi/10.1103/PhysRevLett.128.080502}.

\bibitem[{\citenamefont{Wang and Schirmer}(2010)}]{Schirmer2010}
\bibinfo{author}{\bibfnamefont{X.}~\bibnamefont{Wang}} \bibnamefont{and}
  \bibinfo{author}{\bibfnamefont{S.~G.} \bibnamefont{Schirmer}},
  \emph{\bibinfo{title}{Generating maximal entanglement between non-interacting
  atoms by collective decay and symmetry breaking}} (\bibinfo{year}{2010}),
  \urlprefix\url{https://arxiv.org/abs/1005.2114}.

\bibitem[{\citenamefont{Motzoi et~al.}(2016)\citenamefont{Motzoi, Halperin,
  Wang, Whaley, and Schirmer}}]{Motzoi2016}
\bibinfo{author}{\bibfnamefont{F.}~\bibnamefont{Motzoi}},
  \bibinfo{author}{\bibfnamefont{E.}~\bibnamefont{Halperin}},
  \bibinfo{author}{\bibfnamefont{X.}~\bibnamefont{Wang}},
  \bibinfo{author}{\bibfnamefont{K.~B.} \bibnamefont{Whaley}},
  \bibnamefont{and} \bibinfo{author}{\bibfnamefont{S.}~\bibnamefont{Schirmer}},
  \bibinfo{journal}{Phys. Rev. A} \textbf{\bibinfo{volume}{94}},
  \bibinfo{pages}{032313} (\bibinfo{year}{2016}),
  \urlprefix\url{https://link.aps.org/doi/10.1103/PhysRevA.94.032313}.

\bibitem[{\citenamefont{Reiter et~al.}(2012)\citenamefont{Reiter, Kastoryano,
  and Sørensen}}]{Reiter2012}
\bibinfo{author}{\bibfnamefont{F.}~\bibnamefont{Reiter}},
  \bibinfo{author}{\bibfnamefont{M.~J.} \bibnamefont{Kastoryano}},
  \bibnamefont{and} \bibinfo{author}{\bibfnamefont{A.~S.}
  \bibnamefont{Sørensen}}, \bibinfo{journal}{New Journal of Physics}
  \textbf{\bibinfo{volume}{14}}, \bibinfo{pages}{053022}
  (\bibinfo{year}{2012}),
  \urlprefix\url{https://dx.doi.org/10.1088/1367-2630/14/5/053022}.

\bibitem[{\citenamefont{Govia et~al.}(2022)\citenamefont{Govia, Lingenfelter,
  and Clerk}}]{Govia2022}
\bibinfo{author}{\bibfnamefont{L.~C.~G.} \bibnamefont{Govia}},
  \bibinfo{author}{\bibfnamefont{A.}~\bibnamefont{Lingenfelter}},
  \bibnamefont{and} \bibinfo{author}{\bibfnamefont{A.~A.} \bibnamefont{Clerk}},
  \bibinfo{journal}{Phys. Rev. Res.} \textbf{\bibinfo{volume}{4}},
  \bibinfo{pages}{023010} (\bibinfo{year}{2022}),
  \urlprefix\url{https://link.aps.org/doi/10.1103/PhysRevResearch.4.023010}.

\bibitem[{\citenamefont{Cole et~al.}(2021)\citenamefont{Cole, Wu, Erickson,
  Hou, Wilson, Leibfried, and Reiter}}]{Cole2021}
\bibinfo{author}{\bibfnamefont{D.~C.} \bibnamefont{Cole}},
  \bibinfo{author}{\bibfnamefont{J.~J.} \bibnamefont{Wu}},
  \bibinfo{author}{\bibfnamefont{S.~D.} \bibnamefont{Erickson}},
  \bibinfo{author}{\bibfnamefont{P.-Y.} \bibnamefont{Hou}},
  \bibinfo{author}{\bibfnamefont{A.~C.} \bibnamefont{Wilson}},
  \bibinfo{author}{\bibfnamefont{D.}~\bibnamefont{Leibfried}},
  \bibnamefont{and} \bibinfo{author}{\bibfnamefont{F.}~\bibnamefont{Reiter}},
  \bibinfo{journal}{New Journal of Physics} \textbf{\bibinfo{volume}{23}},
  \bibinfo{pages}{073001} (\bibinfo{year}{2021}),
  \urlprefix\url{https://dx.doi.org/10.1088/1367-2630/ac09c8}.

\bibitem[{\citenamefont{Mamaev et~al.}(2018)\citenamefont{Mamaev, Govia, and
  Clerk}}]{Clerk2018}
\bibinfo{author}{\bibfnamefont{M.}~\bibnamefont{Mamaev}},
  \bibinfo{author}{\bibfnamefont{L.~C.~G.} \bibnamefont{Govia}},
  \bibnamefont{and} \bibinfo{author}{\bibfnamefont{A.~A.} \bibnamefont{Clerk}},
  \bibinfo{journal}{{Quantum}} \textbf{\bibinfo{volume}{2}},
  \bibinfo{pages}{58} (\bibinfo{year}{2018}), ISSN \bibinfo{issn}{2521-327X},
  \urlprefix\url{https://doi.org/10.22331/q-2018-03-27-58}.

\bibitem[{\citenamefont{Roy et~al.}(2015)\citenamefont{Roy, Leghtas, Stone,
  Devoret, and Mirrahimi}}]{Mirrahimi2015}
\bibinfo{author}{\bibfnamefont{A.}~\bibnamefont{Roy}},
  \bibinfo{author}{\bibfnamefont{Z.}~\bibnamefont{Leghtas}},
  \bibinfo{author}{\bibfnamefont{A.~D.} \bibnamefont{Stone}},
  \bibinfo{author}{\bibfnamefont{M.}~\bibnamefont{Devoret}}, \bibnamefont{and}
  \bibinfo{author}{\bibfnamefont{M.}~\bibnamefont{Mirrahimi}},
  \bibinfo{journal}{Phys. Rev. A} \textbf{\bibinfo{volume}{91}},
  \bibinfo{pages}{013810} (\bibinfo{year}{2015}),
  \urlprefix\url{https://link.aps.org/doi/10.1103/PhysRevA.91.013810}.

\bibitem[{\citenamefont{Putterman et~al.}(2022)\citenamefont{Putterman,
  Iverson, Xu, Jiang, Painter, Brand\~ao, and Noh}}]{Noh2022}
\bibinfo{author}{\bibfnamefont{H.}~\bibnamefont{Putterman}},
  \bibinfo{author}{\bibfnamefont{J.}~\bibnamefont{Iverson}},
  \bibinfo{author}{\bibfnamefont{Q.}~\bibnamefont{Xu}},
  \bibinfo{author}{\bibfnamefont{L.}~\bibnamefont{Jiang}},
  \bibinfo{author}{\bibfnamefont{O.}~\bibnamefont{Painter}},
  \bibinfo{author}{\bibfnamefont{F.~G. S.~L.} \bibnamefont{Brand\~ao}},
  \bibnamefont{and} \bibinfo{author}{\bibfnamefont{K.}~\bibnamefont{Noh}},
  \bibinfo{journal}{Phys. Rev. Lett.} \textbf{\bibinfo{volume}{128}},
  \bibinfo{pages}{110502} (\bibinfo{year}{2022}),
  \urlprefix\url{https://link.aps.org/doi/10.1103/PhysRevLett.128.110502}.

\bibitem[{\citenamefont{Leghtas et~al.}(2015)\citenamefont{Leghtas, Touzard,
  Pop, Kou, Vlastakis, Petrenko, Sliwa, Narla, Shankar, Hatridge
  et~al.}}]{Devoret2015}
\bibinfo{author}{\bibfnamefont{Z.}~\bibnamefont{Leghtas}},
  \bibinfo{author}{\bibfnamefont{S.}~\bibnamefont{Touzard}},
  \bibinfo{author}{\bibfnamefont{I.~M.} \bibnamefont{Pop}},
  \bibinfo{author}{\bibfnamefont{A.}~\bibnamefont{Kou}},
  \bibinfo{author}{\bibfnamefont{B.}~\bibnamefont{Vlastakis}},
  \bibinfo{author}{\bibfnamefont{A.}~\bibnamefont{Petrenko}},
  \bibinfo{author}{\bibfnamefont{K.~M.} \bibnamefont{Sliwa}},
  \bibinfo{author}{\bibfnamefont{A.}~\bibnamefont{Narla}},
  \bibinfo{author}{\bibfnamefont{S.}~\bibnamefont{Shankar}},
  \bibinfo{author}{\bibfnamefont{M.~J.} \bibnamefont{Hatridge}},
  \bibnamefont{et~al.}, \bibinfo{journal}{Science}
  \textbf{\bibinfo{volume}{347}}, \bibinfo{pages}{853} (\bibinfo{year}{2015}),
  \urlprefix\url{https://www.science.org/doi/abs/10.1126/science.aaa2085}.

\bibitem[{\citenamefont{Mirrahimi et~al.}(2014)\citenamefont{Mirrahimi,
  Leghtas, Albert, Touzard, Schoelkopf, Jiang, and Devoret}}]{Mirrahimi2014}
\bibinfo{author}{\bibfnamefont{M.}~\bibnamefont{Mirrahimi}},
  \bibinfo{author}{\bibfnamefont{Z.}~\bibnamefont{Leghtas}},
  \bibinfo{author}{\bibfnamefont{V.~V.} \bibnamefont{Albert}},
  \bibinfo{author}{\bibfnamefont{S.}~\bibnamefont{Touzard}},
  \bibinfo{author}{\bibfnamefont{R.~J.} \bibnamefont{Schoelkopf}},
  \bibinfo{author}{\bibfnamefont{L.}~\bibnamefont{Jiang}}, \bibnamefont{and}
  \bibinfo{author}{\bibfnamefont{M.~H.} \bibnamefont{Devoret}},
  \bibinfo{journal}{New Journal of Physics} \textbf{\bibinfo{volume}{16}},
  \bibinfo{pages}{045014} (\bibinfo{year}{2014}),
  \urlprefix\url{https://dx.doi.org/10.1088/1367-2630/16/4/045014}.

\bibitem[{\citenamefont{Reiter et~al.}(2016)\citenamefont{Reiter, Reeb, and
  S\o{}rensen}}]{Sorensen16}
\bibinfo{author}{\bibfnamefont{F.}~\bibnamefont{Reiter}},
  \bibinfo{author}{\bibfnamefont{D.}~\bibnamefont{Reeb}}, \bibnamefont{and}
  \bibinfo{author}{\bibfnamefont{A.~S.} \bibnamefont{S\o{}rensen}},
  \bibinfo{journal}{Phys. Rev. Lett.} \textbf{\bibinfo{volume}{117}},
  \bibinfo{pages}{040501} (\bibinfo{year}{2016}),
  \urlprefix\url{https://link.aps.org/doi/10.1103/PhysRevLett.117.040501}.

\bibitem[{\citenamefont{Aron et~al.}(2016)\citenamefont{Aron, Kulkarni, and
  T\"ureci}}]{Tureci16}
\bibinfo{author}{\bibfnamefont{C.}~\bibnamefont{Aron}},
  \bibinfo{author}{\bibfnamefont{M.}~\bibnamefont{Kulkarni}}, \bibnamefont{and}
  \bibinfo{author}{\bibfnamefont{H.~E.} \bibnamefont{T\"ureci}},
  \bibinfo{journal}{Phys. Rev. X} \textbf{\bibinfo{volume}{6}},
  \bibinfo{pages}{011032} (\bibinfo{year}{2016}),
  \urlprefix\url{https://link.aps.org/doi/10.1103/PhysRevX.6.011032}.

\bibitem[{\citenamefont{Zhou et~al.}(2021)\citenamefont{Zhou, Choi, and
  Lukin}}]{Lukin2021}
\bibinfo{author}{\bibfnamefont{L.}~\bibnamefont{Zhou}},
  \bibinfo{author}{\bibfnamefont{S.}~\bibnamefont{Choi}}, \bibnamefont{and}
  \bibinfo{author}{\bibfnamefont{M.~D.} \bibnamefont{Lukin}},
  \bibinfo{journal}{Phys. Rev. A} \textbf{\bibinfo{volume}{104}},
  \bibinfo{pages}{032418} (\bibinfo{year}{2021}),
  \urlprefix\url{https://link.aps.org/doi/10.1103/PhysRevA.104.032418}.

\bibitem[{\citenamefont{Sweke et~al.}(2013)\citenamefont{Sweke, Sinayskiy, and
  Petruccione}}]{Petruccione2013}
\bibinfo{author}{\bibfnamefont{R.}~\bibnamefont{Sweke}},
  \bibinfo{author}{\bibfnamefont{I.}~\bibnamefont{Sinayskiy}},
  \bibnamefont{and}
  \bibinfo{author}{\bibfnamefont{F.}~\bibnamefont{Petruccione}},
  \bibinfo{journal}{Phys. Rev. A} \textbf{\bibinfo{volume}{87}},
  \bibinfo{pages}{042323} (\bibinfo{year}{2013}),
  \urlprefix\url{https://link.aps.org/doi/10.1103/PhysRevA.87.042323}.

\bibitem[{\citenamefont{Cruz et~al.}(2019)\citenamefont{Cruz, Fournier,
  Gremion, Jeannerot, Komagata, Tosic, Thiesbrummel, Chan, Macris, Dupertuis
  et~al.}}]{Clement2019}
\bibinfo{author}{\bibfnamefont{D.}~\bibnamefont{Cruz}},
  \bibinfo{author}{\bibfnamefont{R.}~\bibnamefont{Fournier}},
  \bibinfo{author}{\bibfnamefont{F.}~\bibnamefont{Gremion}},
  \bibinfo{author}{\bibfnamefont{A.}~\bibnamefont{Jeannerot}},
  \bibinfo{author}{\bibfnamefont{K.}~\bibnamefont{Komagata}},
  \bibinfo{author}{\bibfnamefont{T.}~\bibnamefont{Tosic}},
  \bibinfo{author}{\bibfnamefont{J.}~\bibnamefont{Thiesbrummel}},
  \bibinfo{author}{\bibfnamefont{C.~L.} \bibnamefont{Chan}},
  \bibinfo{author}{\bibfnamefont{N.}~\bibnamefont{Macris}},
  \bibinfo{author}{\bibfnamefont{M.-A.} \bibnamefont{Dupertuis}},
  \bibnamefont{et~al.}, \bibinfo{journal}{Advanced Quantum Technologies}
  \textbf{\bibinfo{volume}{2}}, \bibinfo{pages}{1900015}
  (\bibinfo{year}{2019}),
  \urlprefix\url{https://onlinelibrary.wiley.com/doi/abs/10.1002/qute.201900015}.

\bibitem[{\citenamefont{Mooney et~al.}(2021{\natexlab{a}})\citenamefont{Mooney,
  White, Hill, and Hollenberg}}]{Lloyd2021a}
\bibinfo{author}{\bibfnamefont{G.~J.} \bibnamefont{Mooney}},
  \bibinfo{author}{\bibfnamefont{G.~A.~L.} \bibnamefont{White}},
  \bibinfo{author}{\bibfnamefont{C.~D.} \bibnamefont{Hill}}, \bibnamefont{and}
  \bibinfo{author}{\bibfnamefont{L.~C.~L.} \bibnamefont{Hollenberg}},
  \bibinfo{journal}{Journal of Physics Communications}
  \textbf{\bibinfo{volume}{5}}, \bibinfo{pages}{095004}
  (\bibinfo{year}{2021}{\natexlab{a}}),
  \urlprefix\url{https://dx.doi.org/10.1088/2399-6528/ac1df7}.

\bibitem[{\citenamefont{Mooney et~al.}(2021{\natexlab{b}})\citenamefont{Mooney,
  White, Hill, and Hollenberg}}]{Lloyd2021b}
\bibinfo{author}{\bibfnamefont{G.~J.} \bibnamefont{Mooney}},
  \bibinfo{author}{\bibfnamefont{G.~A.~L.} \bibnamefont{White}},
  \bibinfo{author}{\bibfnamefont{C.~D.} \bibnamefont{Hill}}, \bibnamefont{and}
  \bibinfo{author}{\bibfnamefont{L.~C.~L.} \bibnamefont{Hollenberg}},
  \bibinfo{journal}{Advanced Quantum Technologies}
  \textbf{\bibinfo{volume}{4}}, \bibinfo{pages}{2100061}
  (\bibinfo{year}{2021}{\natexlab{b}}),
  \urlprefix\url{https://onlinelibrary.wiley.com/doi/abs/10.1002/qute.202100061}.

\bibitem[{\citenamefont{Ouyang et~al.}(2022)\citenamefont{Ouyang, Shettell, and
  Markham}}]{Ouyang2022}
\bibinfo{author}{\bibfnamefont{Y.}~\bibnamefont{Ouyang}},
  \bibinfo{author}{\bibfnamefont{N.}~\bibnamefont{Shettell}}, \bibnamefont{and}
  \bibinfo{author}{\bibfnamefont{D.}~\bibnamefont{Markham}},
  \bibinfo{journal}{IEEE Transactions on Information Theory}
  \textbf{\bibinfo{volume}{68}}, \bibinfo{pages}{1809} (\bibinfo{year}{2022}).

\bibitem[{\citenamefont{Prevedel et~al.}(2009)\citenamefont{Prevedel,
  Cronenberg, Tame, Paternostro, Walther, Kim, and Zeilinger}}]{Prevdel2009}
\bibinfo{author}{\bibfnamefont{R.}~\bibnamefont{Prevedel}},
  \bibinfo{author}{\bibfnamefont{G.}~\bibnamefont{Cronenberg}},
  \bibinfo{author}{\bibfnamefont{M.~S.} \bibnamefont{Tame}},
  \bibinfo{author}{\bibfnamefont{M.}~\bibnamefont{Paternostro}},
  \bibinfo{author}{\bibfnamefont{P.}~\bibnamefont{Walther}},
  \bibinfo{author}{\bibfnamefont{M.~S.} \bibnamefont{Kim}}, \bibnamefont{and}
  \bibinfo{author}{\bibfnamefont{A.}~\bibnamefont{Zeilinger}},
  \bibinfo{journal}{Phys. Rev. Lett.} \textbf{\bibinfo{volume}{103}},
  \bibinfo{pages}{020503} (\bibinfo{year}{2009}),
  \urlprefix\url{https://link.aps.org/doi/10.1103/PhysRevLett.103.020503}.

\bibitem[{\citenamefont{Hadfield et~al.}(2019)\citenamefont{Hadfield, Wang,
  O’Gorman, Rieffel, Venturelli, and Biswas}}]{Hadfield2019}
\bibinfo{author}{\bibfnamefont{S.}~\bibnamefont{Hadfield}},
  \bibinfo{author}{\bibfnamefont{Z.}~\bibnamefont{Wang}},
  \bibinfo{author}{\bibfnamefont{B.}~\bibnamefont{O’Gorman}},
  \bibinfo{author}{\bibfnamefont{E.~G.} \bibnamefont{Rieffel}},
  \bibinfo{author}{\bibfnamefont{D.}~\bibnamefont{Venturelli}},
  \bibnamefont{and} \bibinfo{author}{\bibfnamefont{R.}~\bibnamefont{Biswas}},
  \bibinfo{journal}{Algorithms} \textbf{\bibinfo{volume}{12}}
  (\bibinfo{year}{2019}), ISSN \bibinfo{issn}{1999-4893},
  \urlprefix\url{https://www.mdpi.com/1999-4893/12/2/34}.

\bibitem[{\citenamefont{Stockton et~al.}(2003)\citenamefont{Stockton, Geremia,
  Doherty, and Mabuchi}}]{Stockton2003}
\bibinfo{author}{\bibfnamefont{J.~K.} \bibnamefont{Stockton}},
  \bibinfo{author}{\bibfnamefont{J.~M.} \bibnamefont{Geremia}},
  \bibinfo{author}{\bibfnamefont{A.~C.} \bibnamefont{Doherty}},
  \bibnamefont{and} \bibinfo{author}{\bibfnamefont{H.}~\bibnamefont{Mabuchi}},
  \bibinfo{journal}{Phys. Rev. A} \textbf{\bibinfo{volume}{67}},
  \bibinfo{pages}{022112} (\bibinfo{year}{2003}),
  \urlprefix\url{https://link.aps.org/doi/10.1103/PhysRevA.67.022112}.

\bibitem[{\citenamefont{Barnea et~al.}(2015)\citenamefont{Barnea, P\"utz,
  Brask, Brunner, Gisin, and Liang}}]{Barnes2015}
\bibinfo{author}{\bibfnamefont{T.~J.} \bibnamefont{Barnea}},
  \bibinfo{author}{\bibfnamefont{G.}~\bibnamefont{P\"utz}},
  \bibinfo{author}{\bibfnamefont{J.~B.} \bibnamefont{Brask}},
  \bibinfo{author}{\bibfnamefont{N.}~\bibnamefont{Brunner}},
  \bibinfo{author}{\bibfnamefont{N.}~\bibnamefont{Gisin}}, \bibnamefont{and}
  \bibinfo{author}{\bibfnamefont{Y.-C.} \bibnamefont{Liang}},
  \bibinfo{journal}{Phys. Rev. A} \textbf{\bibinfo{volume}{91}},
  \bibinfo{pages}{032108} (\bibinfo{year}{2015}),
  \urlprefix\url{https://link.aps.org/doi/10.1103/PhysRevA.91.032108}.

\bibitem[{\citenamefont{Sen(De) et~al.}(2003)\citenamefont{Sen(De), Sen,
  Wie\ifmmode~\acute{s}\else \'{s}\fi{}niak, Kaszlikowski, and
  \ifmmode~\dot{Z}\else \.{Z}\fi{}ukowski}}]{Sen2003}
\bibinfo{author}{\bibfnamefont{A.}~\bibnamefont{Sen(De)}},
  \bibinfo{author}{\bibfnamefont{U.}~\bibnamefont{Sen}},
  \bibinfo{author}{\bibfnamefont{M.}~\bibnamefont{Wie\ifmmode~\acute{s}\else
  \'{s}\fi{}niak}},
  \bibinfo{author}{\bibfnamefont{D.}~\bibnamefont{Kaszlikowski}},
  \bibnamefont{and}
  \bibinfo{author}{\bibfnamefont{M.}~\bibnamefont{\ifmmode~\dot{Z}\else
  \.{Z}\fi{}ukowski}}, \bibinfo{journal}{Phys. Rev. A}
  \textbf{\bibinfo{volume}{68}}, \bibinfo{pages}{062306}
  (\bibinfo{year}{2003}),
  \urlprefix\url{https://link.aps.org/doi/10.1103/PhysRevA.68.062306}.

\bibitem[{\citenamefont{Ticozzi and Viola}(2014)}]{Ticozzi2014}
\bibinfo{author}{\bibfnamefont{F.}~\bibnamefont{Ticozzi}} \bibnamefont{and}
  \bibinfo{author}{\bibfnamefont{L.}~\bibnamefont{Viola}},
  \bibinfo{journal}{CQuantum Information and Computation}
  \textbf{\bibinfo{volume}{14}}, \bibinfo{pages}{0265} (\bibinfo{year}{2014}).

\bibitem[{\citenamefont{Carmichael}(2009)}]{Carmichael2}
\bibinfo{author}{\bibfnamefont{H.}~\bibnamefont{Carmichael}},
  \emph{\bibinfo{title}{Statistical Methods in Quantum Optics 2: Non-Classical
  Fields}}, Theoretical and Mathematical Physics (\bibinfo{publisher}{Springer
  Berlin Heidelberg}, \bibinfo{year}{2009}), ISBN
  \bibinfo{isbn}{9783540713203}.

\bibitem[{\citenamefont{Kraus et~al.}(2008)\citenamefont{Kraus, B\"uchler,
  Diehl, Kantian, Micheli, and Zoller}}]{Zoller08}
\bibinfo{author}{\bibfnamefont{B.}~\bibnamefont{Kraus}},
  \bibinfo{author}{\bibfnamefont{H.~P.} \bibnamefont{B\"uchler}},
  \bibinfo{author}{\bibfnamefont{S.}~\bibnamefont{Diehl}},
  \bibinfo{author}{\bibfnamefont{A.}~\bibnamefont{Kantian}},
  \bibinfo{author}{\bibfnamefont{A.}~\bibnamefont{Micheli}}, \bibnamefont{and}
  \bibinfo{author}{\bibfnamefont{P.}~\bibnamefont{Zoller}},
  \bibinfo{journal}{Phys. Rev. A} \textbf{\bibinfo{volume}{78}},
  \bibinfo{pages}{042307} (\bibinfo{year}{2008}),
  \urlprefix\url{https://link.aps.org/doi/10.1103/PhysRevA.78.042307}.

\bibitem[{\citenamefont{Leghtas et~al.}(2013)\citenamefont{Leghtas, Vool,
  Shankar, Hatridge, Girvin, Devoret, and Mirrahimi}}]{Leghtas2013}
\bibinfo{author}{\bibfnamefont{Z.}~\bibnamefont{Leghtas}},
  \bibinfo{author}{\bibfnamefont{U.}~\bibnamefont{Vool}},
  \bibinfo{author}{\bibfnamefont{S.}~\bibnamefont{Shankar}},
  \bibinfo{author}{\bibfnamefont{M.}~\bibnamefont{Hatridge}},
  \bibinfo{author}{\bibfnamefont{S.~M.} \bibnamefont{Girvin}},
  \bibinfo{author}{\bibfnamefont{M.~H.} \bibnamefont{Devoret}},
  \bibnamefont{and}
  \bibinfo{author}{\bibfnamefont{M.}~\bibnamefont{Mirrahimi}},
  \bibinfo{journal}{Phys. Rev. A} \textbf{\bibinfo{volume}{88}},
  \bibinfo{pages}{023849} (\bibinfo{year}{2013}),
  \urlprefix\url{https://link.aps.org/doi/10.1103/PhysRevA.88.023849}.

\bibitem[{\citenamefont{Johansson et~al.}(2013)\citenamefont{Johansson, Nation,
  and Nori}}]{QuTiP}
\bibinfo{author}{\bibfnamefont{J.}~\bibnamefont{Johansson}},
  \bibinfo{author}{\bibfnamefont{P.}~\bibnamefont{Nation}}, \bibnamefont{and}
  \bibinfo{author}{\bibfnamefont{F.}~\bibnamefont{Nori}},
  \bibinfo{journal}{Computer Physics Communications}
  \textbf{\bibinfo{volume}{184}}, \bibinfo{pages}{1234} (\bibinfo{year}{2013}),
  ISSN \bibinfo{issn}{0010-4655},
  \urlprefix\url{https://www.sciencedirect.com/science/article/pii/S0010465512003955}.

\bibitem[{\citenamefont{G\"uhne and T\'oth}(2009)}]{Toth2009}
\bibinfo{author}{\bibfnamefont{O.}~\bibnamefont{G\"uhne}} \bibnamefont{and}
  \bibinfo{author}{\bibfnamefont{G.}~\bibnamefont{T\'oth}},
  \bibinfo{journal}{Physics Reports} \textbf{\bibinfo{volume}{474}},
  \bibinfo{pages}{1} (\bibinfo{year}{2009}), ISSN \bibinfo{issn}{0370-1573},
  \urlprefix\url{https://www.sciencedirect.com/science/article/pii/S0370157309000623}.

\bibitem[{\citenamefont{Gu et~al.}(2017)\citenamefont{Gu, Kockum, Miranowicz,
  xi~Liu, and Nori}}]{Gu2017}
\bibinfo{author}{\bibfnamefont{X.}~\bibnamefont{Gu}},
  \bibinfo{author}{\bibfnamefont{A.~F.} \bibnamefont{Kockum}},
  \bibinfo{author}{\bibfnamefont{A.}~\bibnamefont{Miranowicz}},
  \bibinfo{author}{\bibfnamefont{Y.}~\bibnamefont{xi~Liu}}, \bibnamefont{and}
  \bibinfo{author}{\bibfnamefont{F.}~\bibnamefont{Nori}},
  \bibinfo{journal}{Physics Reports} \textbf{\bibinfo{volume}{718-719}},
  \bibinfo{pages}{1} (\bibinfo{year}{2017}), ISSN \bibinfo{issn}{0370-1573},
  \bibinfo{note}{microwave photonics with superconducting quantum circuits},
  \urlprefix\url{https://www.sciencedirect.com/science/article/pii/S0370157317303290}.

\bibitem[{\citenamefont{Leibfried et~al.}(2003)\citenamefont{Leibfried, Blatt,
  Monroe, and Wineland}}]{Leibfried2003}
\bibinfo{author}{\bibfnamefont{D.}~\bibnamefont{Leibfried}},
  \bibinfo{author}{\bibfnamefont{R.}~\bibnamefont{Blatt}},
  \bibinfo{author}{\bibfnamefont{C.}~\bibnamefont{Monroe}}, \bibnamefont{and}
  \bibinfo{author}{\bibfnamefont{D.}~\bibnamefont{Wineland}},
  \bibinfo{journal}{Rev. Mod. Phys.} \textbf{\bibinfo{volume}{75}},
  \bibinfo{pages}{281} (\bibinfo{year}{2003}),
  \urlprefix\url{https://link.aps.org/doi/10.1103/RevModPhys.75.281}.

\bibitem[{\citenamefont{Kaufman et~al.}(2012)\citenamefont{Kaufman, Lester, and
  Regal}}]{Kaufman2012}
\bibinfo{author}{\bibfnamefont{A.~M.} \bibnamefont{Kaufman}},
  \bibinfo{author}{\bibfnamefont{B.~J.} \bibnamefont{Lester}},
  \bibnamefont{and} \bibinfo{author}{\bibfnamefont{C.~A.} \bibnamefont{Regal}},
  \bibinfo{journal}{Phys. Rev. X} \textbf{\bibinfo{volume}{2}},
  \bibinfo{pages}{041014} (\bibinfo{year}{2012}),
  \urlprefix\url{https://link.aps.org/doi/10.1103/PhysRevX.2.041014}.

\bibitem[{\citenamefont{Aspelmeyer et~al.}(2014)\citenamefont{Aspelmeyer,
  Kippenberg, and Marquardt}}]{Aspelmeyer2014}
\bibinfo{author}{\bibfnamefont{M.}~\bibnamefont{Aspelmeyer}},
  \bibinfo{author}{\bibfnamefont{T.~J.} \bibnamefont{Kippenberg}},
  \bibnamefont{and}
  \bibinfo{author}{\bibfnamefont{F.}~\bibnamefont{Marquardt}},
  \bibinfo{journal}{Rev. Mod. Phys.} \textbf{\bibinfo{volume}{86}},
  \bibinfo{pages}{1391} (\bibinfo{year}{2014}),
  \urlprefix\url{https://link.aps.org/doi/10.1103/RevModPhys.86.1391}.

\bibitem[{\citenamefont{Hillmann et~al.}(2020)\citenamefont{Hillmann,
  Quijandr\'{\i}a, Johansson, Ferraro, Gasparinetti, and
  Ferrini}}]{Hillman2020}
\bibinfo{author}{\bibfnamefont{T.}~\bibnamefont{Hillmann}},
  \bibinfo{author}{\bibfnamefont{F.}~\bibnamefont{Quijandr\'{\i}a}},
  \bibinfo{author}{\bibfnamefont{G.}~\bibnamefont{Johansson}},
  \bibinfo{author}{\bibfnamefont{A.}~\bibnamefont{Ferraro}},
  \bibinfo{author}{\bibfnamefont{S.}~\bibnamefont{Gasparinetti}},
  \bibnamefont{and} \bibinfo{author}{\bibfnamefont{G.}~\bibnamefont{Ferrini}},
  \bibinfo{journal}{Phys. Rev. Lett.} \textbf{\bibinfo{volume}{125}},
  \bibinfo{pages}{160501} (\bibinfo{year}{2020}),
  \urlprefix\url{https://link.aps.org/doi/10.1103/PhysRevLett.125.160501}.

\bibitem[{\citenamefont{Ye et~al.}(2021)\citenamefont{Ye, Peng, Naghiloo,
  Cunningham, and O'Brien}}]{Ye2021}
\bibinfo{author}{\bibfnamefont{Y.}~\bibnamefont{Ye}},
  \bibinfo{author}{\bibfnamefont{K.}~\bibnamefont{Peng}},
  \bibinfo{author}{\bibfnamefont{M.}~\bibnamefont{Naghiloo}},
  \bibinfo{author}{\bibfnamefont{G.}~\bibnamefont{Cunningham}},
  \bibnamefont{and} \bibinfo{author}{\bibfnamefont{K.~P.}
  \bibnamefont{O'Brien}}, \bibinfo{journal}{Phys. Rev. Lett.}
  \textbf{\bibinfo{volume}{127}}, \bibinfo{pages}{050502}
  (\bibinfo{year}{2021}),
  \urlprefix\url{https://link.aps.org/doi/10.1103/PhysRevLett.127.050502}.

\bibitem[{\citenamefont{Menke et~al.}(2022)\citenamefont{Menke, Banner,
  Bergamaschi, Di~Paolo, Veps\"al\"ainen, Weber, Winik, Melville, Niedzielski,
  Rosenberg et~al.}}]{Menke2022}
\bibinfo{author}{\bibfnamefont{T.}~\bibnamefont{Menke}},
  \bibinfo{author}{\bibfnamefont{W.~P.} \bibnamefont{Banner}},
  \bibinfo{author}{\bibfnamefont{T.~R.} \bibnamefont{Bergamaschi}},
  \bibinfo{author}{\bibfnamefont{A.}~\bibnamefont{Di~Paolo}},
  \bibinfo{author}{\bibfnamefont{A.}~\bibnamefont{Veps\"al\"ainen}},
  \bibinfo{author}{\bibfnamefont{S.~J.} \bibnamefont{Weber}},
  \bibinfo{author}{\bibfnamefont{R.}~\bibnamefont{Winik}},
  \bibinfo{author}{\bibfnamefont{A.}~\bibnamefont{Melville}},
  \bibinfo{author}{\bibfnamefont{B.~M.} \bibnamefont{Niedzielski}},
  \bibinfo{author}{\bibfnamefont{D.}~\bibnamefont{Rosenberg}},
  \bibnamefont{et~al.}, \bibinfo{journal}{Phys. Rev. Lett.}
  \textbf{\bibinfo{volume}{129}}, \bibinfo{pages}{220501}
  (\bibinfo{year}{2022}),
  \urlprefix\url{https://link.aps.org/doi/10.1103/PhysRevLett.129.220501}.

\bibitem[{\citenamefont{Wang et~al.}(2022)\citenamefont{Wang, Rajabzadeh, Lee,
  and Safavi-Naeini}}]{Wang2022}
\bibinfo{author}{\bibfnamefont{Z.}~\bibnamefont{Wang}},
  \bibinfo{author}{\bibfnamefont{T.}~\bibnamefont{Rajabzadeh}},
  \bibinfo{author}{\bibfnamefont{N.}~\bibnamefont{Lee}}, \bibnamefont{and}
  \bibinfo{author}{\bibfnamefont{A.~H.} \bibnamefont{Safavi-Naeini}},
  \bibinfo{journal}{PRX Quantum} \textbf{\bibinfo{volume}{3}},
  \bibinfo{pages}{020302} (\bibinfo{year}{2022}),
  \urlprefix\url{https://link.aps.org/doi/10.1103/PRXQuantum.3.020302}.

\bibitem[{\citenamefont{Petroff}(2021)}]{Petroff21}
\bibinfo{author}{\bibfnamefont{M.~A.} \bibnamefont{Petroff}},
  \emph{\bibinfo{title}{Accessible color sequences for data visualization}}
  (\bibinfo{year}{2021}), \urlprefix\url{https://arxiv.org/abs/2107.02270}.

\end{thebibliography}
%

%
\end{document}